\documentclass[fleqn,usenatbib]{mnras}

\usepackage{newtxtext,newtxmath}

\usepackage[T1]{fontenc}

\DeclareRobustCommand{\VAN}[3]{#2}
\let\VANthebibliography\thebibliography
\def\thebibliography{\DeclareRobustCommand{\VAN}[3]{##3}\VANthebibliography}

\usepackage{graphicx}	
\usepackage{amsmath}	
\usepackage{amssymb}	
\usepackage{multirow}
\usepackage[table]{xcolor}

\newcommand{\Hcal}{\mathcal{H}}
\newcommand{\dev}[2]{\frac{\text{d} #1}{\text{d} #2}}
\newcommand{\pdev}[2]{\frac{\partial #1}{\partial #2}}
\newcommand{\pdevsec}[2]{\frac{\partial^2 #1}{\partial #2^2}}

\title[Stochastic Accretion Discs]{Investigating the Theory of Propagating Fluctuations with Numerical Models of Stochastic Accretion Discs}

\author[S. G. D. Turner and C. S. Reynolds]{
Samuel G. D. Turner\thanks{sgdt2@cam.ac.uk}
and Christopher S. Reynolds
\\
Institute of Astronomy, University of Cambridge, Madingley Road, Cambridge, CB3 0HA, UK
}

\date{Accepted XXX. Received YYY; in original form ZZZ}

\pubyear{2020}

\begin{document}
\label{firstpage}
\pagerange{\pageref{firstpage}--\pageref{lastpage}}
\maketitle

\begin{abstract}

Across a large range of scales, accreting sources show remarkably similar patterns of variability, most notably the log-normality of the luminosity distribution and the linear root-mean square (rms)-flux relationship. These results are often explained using the theory of propagating fluctuations in which fluctuations in the viscosity create perturbations in the accretion rate at all radii, propagate inwards and combine multiplicatively. While this idea has been extensively studied analytically in a linear regime, there has been relatively little numerical work investigating the non-linear behaviour. In this paper, we present a suite of stochastically driven 1-d $\alpha$-disc simulations, exploring the behaviour of these discs. We find that the eponymous propagating fluctuations are present in all simulations across a wide range of model parameters, in contradiction to previous work. Of the model parameters, we find by far the most important to be the timescale on which the viscosity fluctuations occur. Physically, this timescale will depend on the underlying physical mechanism, thought to be the magnetorotational instability (MRI). We find a close relationship between this fluctuation timescale and the break frequency in the power spectral density (PSD) of the luminosity, a fact which could allow observational probes of the behaviour of the MRI dynamo. We report a fitting formula for the break frequency as a function of the fluctuation timescale, the disc thickness and the mass of the central object.

\end{abstract}




\graphicspath{{Figures/}}

\section{Introduction}

{ From protostars to active galactic nuclei (AGN) accretion discs show remarkable similarity in their observed variability properties despite the vast range of scales and diversity of physics involved.  Broadly,} the power spectral densities (PSDs) of {accreting objects reveal} variability in their luminosities {that} span many decades in temporal frequency. Specifically, variability is observed on timescales much longer that the timescale of any physical process within the {central regions which dominate the observed emissions.}  The luminosity {fluctuations are found to be} log-normally distributed and there is a linear relationship between the root-mean square (rms) variation on short timescales and the mean flux varying over longer timescales (e.g. \citealt{Uttley&McHardy2001}). {\citet{Uttley+2005} showed that if this linear rms-flux relationship extends to all timescales (as suggested by observational data), then the corresponding light curve must be log-normally distributed.} {These properties have} been observed in a wide range of {accreting} sources including AGN in X-ray \citep{Gaskell2004, Vaughan+2011} and optical \citep{Lyutyi&Oknyanskii1987}, X-ray binaries (XRBs) also in X-ray \citep{Gleissner+2004} and optical \citep{Gandhi2009}, cataclysmic variables (CVs) \citep{Scaringi+2012}, {and young stellar objects \citep{Scaringi+2015}.}

{Further, }coherence is observed between radiation {emitted from the accretion disc} in different energy bands. The {variability} between two bands is found to be coherent at low {temporal} frequencies and becomes incoherent at high frequencies \citep{Markowitz+2007}. {Associated with the coherence, time} lags are found between the energy bands in {that} the flux received in one band leads that in another. These are termed hard lags if the higher energy radiation trails the lower energy radiation \citep{Nowak2000, Markowitz2005, Arevalo+2006} and soft lags if the lower energy radiation trails \citep{Fabian+2009, DeMarco+2011, Scaringi+2013}. These lags are typically frequency dependent (here meaning temporal frequency not radiation frequency) and some sources show both hard and soft lags between the same energy bands at different frequencies \citep{Fabian+2009, Zoghbi+2010}. {Again, these lags are seen in AGN, X-ray binaries \citep{Nowak2000}, and cataclysmic variables \citep{Scaringi+2013}.}

One of the most successful models for explaining the observed variability from accretion discs is the theory of propagating fluctuations, first described by \citet{Lyubarskii1997}. {The starting point for this model is a radiatively-efficient, geometrically-thin fluid disc that accretes due to angular momentum transport associated with an effective viscosity $\nu$.  \citet{Lyubarskii1997} adopts the standard $\alpha$-prescription \citep[][also see \S\ref{sec:models}]{Shakura&Sunyaev1973} for the effective viscosity, and then proceeds to add small stochastic perturbations driven on the viscous timescale to the value of $\alpha$ at each radius. Linearizing the disc equations with respect to the small stochastic perturbation, \citet{Lyubarskii1997} proceeds to show that the observed disc variability has a flicker noise ($1/f$) PSD, extending down to temporal frequencies much lower than those characterizing the inner region of the disc that dominates the emission.  This is due to the fact that viscosity} perturbations create {fluctuations} in the local accretion rate at all radii {which then} propagate inwards viscously and combine together, creating broadband variability in the accretion rate through the inner regions of the disc. The propagation of low frequency variability {from the outer disc} into the inner region of the disc {that dominates the observed emission} gives a natural explanation for observed variability {at low temporal frequency}. It also gives an explanation for hard lags as the fluctuations first propagate through the cooler, outer regions of the disc before {hitting} the hotter, inner regions.

{The linearized analysis of \citet{Lyubarskii1997} was unable to capture the linear rms-flux relation and log-normal behaviour seen in real systems.  The connection between the propagating fluctuations and these non-linear properties was intuited by \citet{Uttley+2005} on the basis that the fluctuations in mass accretion rate should be multiplicatively modulated by local fluctuations in the viscosity parameter.  

The first steps towards a true non-linear generalization of the \citet{Lyubarskii1997} approach was taken by \citet{Cowperthwaite&Reynolds2014}.  They employed a slightly simpler viscosity prescription, but did not restrict to small stochastic viscosity  perturbations.  Their numerical solution of the disc evolution equation confirmed that propagating fluctuations produced log-normal flux distributions, linear rms-flux relations, and frequency-dependent time-lags \citep[on this last point, also see][]{Ahmad+2018}. \citet{Cowperthwaite&Reynolds2014} suggested that the viscosity fluctuations needed to be driven sufficiently slowly, on the order of the viscous timescale, in order to produce this non-linear variability.

These questions of broad-band disc variability have also been addressed in 3-d magnetohydrodynamic (MHD) simulations. In these models, the (fluctuating) angular momentum transport is a natural consequence of the correlated Reynolds- and Maxwell-stresses in the MHD turbulence that results from the magnetorotational instability \citep{Balbus&Hawley1991}. The high-resolution MHD simulations of a geometrically-thin disc by \citet{Hogg&Reynolds2016} found evidence for propagating fluctuations and recovered the non-linear variability \citep[also see ][]{Bollimpalli+2020}.  A detailed examination of these models revealed that the local angular momentum transport was modulated principally by a quasi-periodic dynamo cycle which emerges from the MHD turbulence (with a characteristic local timescale of approximately 10 orbital periods), rather than the rapid (orbital timescale) turbulent fluctuations themselves. Given that the dynamo cycles are still much faster than the viscous evolution, the existence of the propagating fluctuations and the non-linear luminosity variability in the MHD models appeared to be at odds with the simple viscous models of \citet{Cowperthwaite&Reynolds2014}.

While full MHD simulations remove the need to make any ad-hoc assumptions about the nature or stochasticity of the effective viscosity, their computational expense sets stringent limits on the number of disc orbits that can be followed and hence the range of temporal frequencies that can be explored.  Motivated by this, as well as the potential discrepancy between the MHD results with the findings of \citet{Cowperthwaite&Reynolds2014}, in this paper we return to reduced-dimensionality models employing a stochastic viscosity.  Extending the work of \citet{Cowperthwaite&Reynolds2014}, we construct models based on stochastically-modulated $\alpha$-viscosity, exploring the nature of the observable variability as a function of the assumed stochastic model (functional form and driving timescale), and disc thickness.  We find luminosity PSDs with a broken power-law form, and investigate the dependence of the slopes and break frequency on disc thickness and driving time-scale. We also find that the PDF of the accretion rate across the inner edge of the disc is distinctly different from the luminosity PSDs.  We characterise the coherence and frequency dependent time lags of the the local accretion rate at different radii, finding that the fluctuations are coherent on timescales longer than the viscous time and also time delayed by the viscous travel time.  The local dissipation/emission at different radii is also coherent on timescales longer than the viscous time, but with a time-delay that is appreciably smaller (in agreement with \cite{Ahmad+2018}).  Finally, we recover non-linear variability even for rapidly modulated viscosity, and we reconcile differences with \citet{Cowperthwaite&Reynolds2014} as being due to the form of the viscosity prescription (with the $\alpha$-prescription introducing stronger non-linearities). 

The paper is organized as follows.  Section~2 describes the basic model set-up and the sketches our numerical scheme.  In Section~3 we discuss in detail the variability properties of our fiducial model, and Section~4 explores models with varying stochastic models, disc thickness, and driving timescales.  Section~5 briefly discusses chromatic variability of the disc (using local blackbody emission).  Section~6 presents a discussion of the set of results, before drawing our conclusions in Section~7. 
}

\section{Theory and Method}

\subsection{Theoretical Models}\label{sec:models}

The theoretical and numerical model used in this work is {motivated} by \citet{Cowperthwaite&Reynolds2014}. Differences between our approach and theirs are highlighted where appropriate.

The starting point for our model is the standard 1D disc diffusion equation (e.g. \citealt{Pringle1981})
\begin{equation}
    \label{eq:diffusion_eq}
    \pdev{\Sigma}{t} = \frac{3}{R}\pdev{}{R}\left[\sqrt{R}\pdev{}{R}\left(\nu\Sigma\sqrt{R}\right)\right]\, ,
\end{equation}
where $R$ is the radial distance from the central object, $t$ is time, $\Sigma$ is the surface density of the disc and ${\nu=\eta/\rho}$ is the kinematic viscosity.

This equation is derived assuming that the disc is geometrically `thin', meaning that the radial behaviour is decoupled from the vertical structure. This requires that
\begin{equation}
    \label{eq:aspect_ratio}
    \Hcal \equiv \frac{H}{R} \ll 1 \, ,
\end{equation}
where $H$ is the local scale-height of the disc. It can be shown (e.g. \citealt{Frank+2002}) that for a thin disc $\Hcal$ is given by
\begin{equation}
    \label{eq:aspect_ratio_2}
    \Hcal = \frac{c_s}{v_K} \, ,
\end{equation}
where ${c_s = (P/\rho)^{1/2}}$ is the local isothermal sound speed and $v_K$ is the local Keplerian velocity. Therefore, the requirement that ${\Hcal\ll1}$ is equivalent to requiring that the Keplerian velocity be highly supersonic. {An analysis of the energy equation of viscous hydrodynamics shows that the ratio of the radially-advected thermal energy to the locally dissipated energy is of order ${\Hcal}^2$ \citep{Abramowicz+1995}, showing that these thin discs are also radiatively-efficient.}

{ When it is assumed} that ${\nu\propto R^a}$ for some constant $a$, eq. \eqref{eq:diffusion_eq} can be tackled via a Green's function approach \citep{Lynden-Bell&Pringle1974, Pringle1991, Tanaka2011, Lipunova2015, Balbus2017, Mushtukov+2019}. More general { forms of viscosity }can be tackled via a numerical integration scheme.

In \citet{Cowperthwaite&Reynolds2014} the viscosity was assumed to be of the form ${\nu=\nu_0R}$. In this work we instead use the {more physically-motivated} $\alpha$ model of \citet{Shakura&Sunyaev1973} which states that
\begin{equation}
    \label{eq:alpha_viscosity}
    \nu = \alpha c_sH \, ,
\end{equation}
where $\alpha$ is a numerical parameter. This prescription is motivated by a turbulent viscosity given by ${\nu}=v_\text{turb}\ell_\text{turb}$ where $v_\text{turb}$ and $\ell_\text{turb}$ are the turbulent speed and length scale respectively. It is physically reasonable to expect these scales to be limited by the local sound speed and the disc thickness. It is therefore expected that ${0<\alpha\lesssim1}$ which contains the underlying physics in a simple numerical parameter. This $\alpha$ prescription can also be motivated from a dimensional approach, where $c_s$ and $H$ are the natural velocity and length scales to consider. While apparently simple, it has been shown that the $\alpha$ prescription works well in describing the mean behaviour of MRI turbulence \citep{Balbus&Papaloizou1999}.

While eq. \eqref{eq:diffusion_eq} can be integrated forward in time numerically, we can rewrite it into a more {convenient form} by letting ${x=\sqrt{R/r_g}}$ and ${\tilde{t}=t/t_g}$ where ${r_g=GM_\bullet/c^2}$ and ${t_g=GM_\bullet/c^3}$ are the gravitational radius and gravitational time respectively. Here, $M_\bullet$ is the mass of the central object and $G$ and $c$ are the gravitational constant and speed of light respectively. With this, eq. \eqref{eq:alpha_viscosity} becomes
\begin{equation}
    \label{eq:alpha_viscosity_x}
    \nu = \alpha\Hcal^2x [r_gc] \, .
\end{equation}
This is in contrast to the \citet{Cowperthwaite&Reynolds2014} prescriptions of ${\nu=\nu_0x^2[r_g]}$. A cursory examination of eq. \eqref{eq:alpha_viscosity_x} shows that we have simply shifted the requirement to specify a function of viscosity onto one for $\Hcal$. While it is possible to calculate ${\Hcal=\Hcal(x,\Sigma)}$ at all radii and times (e.g. \citealt{Frank+2002}), in this work we make the simplifying assumption that $\Hcal$ takes a constant value everywhere. This value of $\Hcal$ is one of the input parameters of our model.

Further, {defining} ${\Psi=\Sigma x^2}$ and using eq. \eqref{eq:alpha_viscosity_x} for $\nu$, eq. \eqref{eq:diffusion_eq} becomes
\begin{equation}
    \label{eq:diffusion_eq_x}
    \pdev{\Psi}{\tilde{t}} = \frac{3}{4x} \pdevsec{}{x}\left(\alpha\Hcal^2\Psi\right) \, .
\end{equation}
This equation is the fundamental equation {of our model which we solve numerically} over a grid which is linearly spaced in $x$, which we shall refer to as the diffusion grid.

It is informative to look at the instantaneous accretion rate as a function of radius and time. Under the same assumptions used in deriving eq. \eqref{eq:diffusion_eq}, it can be shown that the radial velocity is given by
\begin{equation}
    \label{eq:thin_radial_velocity}
    v_R(R,t) = -\frac{3}{\Sigma\sqrt{R}}
    \pdev{}{R}\left(\nu\Sigma\sqrt{R}\right)\, ,
\end{equation}
and {so} the local instantaneous mass accretion rate is given by
\begin{equation}
    \label{eq:accretion_rate}
    \dot{M}(R,t) = 6\pi\sqrt{R}\pdev{}{R}\left(\nu\Sigma\sqrt{R}\right)
    = 3\pi\pdev{}{x}\left(\alpha\Hcal^2\Psi\right) [r_gc] \, .
\end{equation}

It can also be shown that the heating rate (from viscous dissipation) per unit surface area of the disc is given by
\begin{equation}
    \label{eq:heating_rate}
    D(R,t) = \frac{9}{8}\nu\Sigma\frac{GM_\bullet}{R^3}
    = \frac{9}{8}\frac{\alpha\Hcal^2\Psi}{x^7} \left[\frac{c^3}{r_g}\right] \, .
\end{equation}
It is worth noting that both ${\dot{M}(x,t)}$ and ${D(x,t)}$ are given as changes per unit real time $t$, not per unit code time $\tilde{t}$.

We can take eq. \eqref{eq:heating_rate} and convert it into a total luminosity from the disc by integrating ${4\pi RD(R)\text{d}R}$ over the entire disc. Note that the integrand is $4\pi$ not $2\pi$ as we need to capture the emission from both the top and bottom surfaces of the disc. This gives
\begin{equation}
    \label{eq:luminosity}
    L = \int\frac{9}{2}\pi\nu\Sigma\frac{GM_\bullet}{R^2}\text{d}R
    = \int9\pi\frac{\alpha\Hcal^2\Psi}{x^4}\text{d}x \left[r_gc^3\right] \, .
\end{equation}

\subsubsection{Stochastic Model for Viscosity}\label{sec:stoch_model}

At the core of our {approach} is the stochastic model for $\alpha$. We assume that $\alpha$ is a function of a stochastic random variable $\beta$ whose temporal variation is given by an Ornstein-Uhlenbeck (OU) process
\begin{equation}
    \label{eq:beta}
    \text{d}\beta(t) = -\omega_0(\beta(t)-\mu)\text{d}t + \xi\text{d}W(t) \, .
\end{equation}
Here, $\omega_0$ is the characteristic frequency of the process, $\mu$ is the mean value of $\beta$, $\xi$ is the amplitude of the driving process and $\text{d}W$ is Gaussian white noise, the derivative of a Wiener process. The OU process means that $\beta$ is normally distributed about $\mu$ when considering sufficiently long timescales (compared to ${1/\omega_0}$). Note that although this is written in terms of $t$, when implemented in the model this is calculated in $\tilde{t}$ rather than $t$, with ${\tilde{\omega}_0\sim1/\tilde{t}}$.

For this work, we require that $\mu=0$ and so $\beta$ always has a mean $0$. If $\xi=0$, eq. \eqref{eq:beta} shows that $\beta$ will follow an exponential decay with an e-folding time of $1/\omega_0$. Therefore, $\omega_0$ can be thought of as a decay frequency in the absence of a driving term.

Requiring that $\text{d}W{=}\mathcal{N}(0,1)$, it can be shown that $\xi$ can be written in terms of the rms value of $\beta$ as \citep{Kelly+2011}
\begin{equation}
    \label{eq:rmsbeta}
    \xi =  \sqrt{2\omega_0\left<\beta^2\right>\text{d}t} \, .
\end{equation}
This allows the value of $\sqrt{\left<\beta^2\right>}$ to be specified instead of $\xi$, which is a more natural variable to work with.

At this point, we must consider the radial nature of $\beta$. In previous work \citet{Cowperthwaite&Reynolds2014} specified an independent value of $\beta$ in each grid cell. Each of these values evolved according to eq. \eqref{eq:beta} where $\omega_0$ (and so $\xi$) varied with radius. One potential issue with this approach is that it essentially sets the coherence length for the viscosity to be the grid spacing. One possible result of this is that it could make the model output resolution dependent.

In this work, we set-up the $\beta$ values to be coherent on a certain length scale. We set this length scale to be the local disc height and so require ${\Delta R = H}$ where $\Delta R$ is the coherence length. Recasting this in terms of our radial variable $x$ gives
\begin{equation}
    \label{eq:coherence_length}
    \Delta x =  \frac{1}{2}x\Hcal \, .
\end{equation}
Using this equation, we create a discrete grid ({distinct from} the diffusion grid for solving eq. \ref{eq:diffusion_eq_x}) of for the evolution of beta, with the location of the grid cells given by
\begin{equation}
    \label{eq:coherence_recurrence}
    x_i = x_{i-1}+\Delta x_{i-1} = x_{i-1}\left(1+\frac{1}{2}\Hcal\right)
    = x_0\left(1+\frac{1}{2}\Hcal\right)^i \, ,
\end{equation}
where the innermost grid cell given by ${x_0=x_\text{in}}$. We call this grid the $\beta$ grid. These $\beta$ cells evolve independently with time according to eq. \eqref{eq:beta}.

When numerically solving eq. \eqref{eq:diffusion_eq_x}, we require the value of $\beta$ at the locations of the diffusion grid. To find these values, we {perform a second-order interpolation} between the values of $\beta$ specified on the $\beta$ grid, giving a continuous function $\beta(x,t)$. To ensure that the radial variation in $\beta$ is adequately captured, we require that the diffusion grid be everywhere finer than the $\beta$ grid.

We consider two different models for $\alpha(\beta)$. The first follows \citet{Cowperthwaite&Reynolds2014} and defines
\begin{equation}
    \label{eq:linear_alpha}
    \alpha = \alpha_0(1+\beta) \, ,
\end{equation}
where $\alpha_0$ is a parameter of the model. We will refer to this as the `linear' model for $\alpha$.  Eq. \eqref{eq:linear_alpha} creates a normally distributed $\alpha$.

While the linear model seems the most natural to use, it runs into a potentially serious issue; should $\alpha$ become negative in any part of the disc, eq. \eqref{eq:diffusion_eq_x} essentially becomes an anti-diffusion equation where variations in the surface density grow rather than being smoothed out, which presents a stability issue for our simple 1D scheme. To avoid this, we introduce a floor when calculating $\alpha$ which restricts it to being non-negative. This floor is unphysical and could potentially affect the results of the simulations. How large an effect it has will depend on how often the floor is required, which in turn will depend on the value of $\sqrt{\left<\beta^2\right>}$.

We also consider an `exponential' model defined by
\begin{equation}
    \label{eq:exponential_alpha}
    \alpha = \alpha_0e^{\beta} \, .
\end{equation}
We can see that this model does not require a floor since ${\alpha>0}$ by definition. Another advantage of this prescription is that the log-normality of $\alpha$ is consistent with the distribution of the effective $\alpha$ found by \citet{Hogg&Reynolds2016} using a full MHD simulation.

Thus far we have not specified the form of $\omega_0(x)$. We will specify this in terms of its reciprocal, which we call the driving timescale. The standard assumption (following \citealt{Lyubarskii1997}) to make here is that the driving timescale is equal to the viscous timescale, given by
\begin{equation}
    \label{eq:t_visc_g}
    t_{\nu,\text{g}} = \frac{R^2}{\nu}
    \implies
    \tilde{t}_{\nu,\text{g}} = \frac{x^3}{\alpha_0\Hcal^2} \, ,
\end{equation}
where $\alpha_0$ is the unperturbed value of $\alpha$. In deriving this expression it was assumed that the relevant length scale for viscous diffusion is $R$. For this reason we refer to this as the global viscous timescale, signified by the subscript `g'. In eq. \eqref{eq:coherence_length} we defined the coherence length for the viscosity. This gives a second viscous timescale of
\begin{equation}
    \label{eq:t_visc_c}
    t_{\nu,\text{c}} = \frac{\Delta R^2}{\nu} = \frac{\Hcal^2R^2}{\nu}
    \implies
    \tilde{t}_{\nu,\text{c}} = \frac{x^3}{\alpha_0} \, ,
\end{equation}
which we shall refer to as the coherence length viscous timescale, signified by the subscript `c'.  {We note that $t_{\nu,\text{c}}$ is the same as the classical thermal timescale and, for $\alpha\sim 0.1$, is of the same order as the local dynamo-cycle period found in MHD simulations \citep{Hogg&Reynolds2018}.}

Similarly, the orbital timescale can be defined as the reciprocal of the orbital angular velocity, $\Omega$, as
\begin{equation}
    \label{eq:t_orbital}
    t_\phi = \frac{1}{\Omega} = \left(\frac{R^3}{GM_\bullet}\right)^{1/2}
    \implies
    \tilde{t}_\phi = x^3 \, .
\end{equation}
We can see that, under our assumption of constant $\Hcal$, these three timescales are simply linear multiples of each other.

In our model, we can use any of these timescales (and multiples thereof) in defining $\tilde{\omega}_0\sim1/\tilde{t}$. The values of $\tilde{\omega}_0(x)$ are calculated based on the initial conditions and do not vary as a function of time. This reflects the fact that this timescale is a useful parameterisation of the timescale on which the viscosity varies, and may not reflect the underlying physical model.

\subsubsection{Steady State}

If the assumption of steady state is made, then we can recover an analytic solution for the state of the disc. In steady state, the accretion rate $\dot{M}$ becomes a constant not just in time but also radius. If it were not then a difference in accretion rate between different radii would lead to a pile-up of material, thus violating the steady state assumption.

In order to derive the steady-state solution, one further assumption is required about a boundary condition for the disc. The standard boundary condition used is that of a zero-torque boundary at the inner edge of the disc. This is a good assumption not just for black hole discs where the inner edge occurs at $r_\text{ISCO}$, the innermost stable circular orbit, but also for accretion onto stars which feature a boundary layer where the disc meets the surface of the star. With this zero-torque condition, the steady state surface density is given by
\begin{equation}
    \label{eq:Sigma_SS}
    \Sigma = \frac{\dot{M}}{3\pi\nu}\left(1 - \sqrt{\frac{R_*}{R}}\right)
    = \frac{\dot{M}}{3\pi\alpha\Hcal^2x^2}\left(x - x_*\right)\left[\frac{1}{r_gc}\right] \, ,
\end{equation}
which becomes
\begin{equation}
    \label{eq:Psi_SS}
    \Psi = \frac{\dot{M}}{3\pi\alpha\Hcal^2}\left(x - x_*\right)\left[\frac{1}{r_gc}\right] \, .
\end{equation}

\subsection{{Numerical Scheme}}

All simulations are performed using an explicit numerical scheme. The basic grid on which the diffusion equation is evolved extends from ${x=\sqrt{6}}$ to ${x=100}$. This inner boundary is chosen as ${R=6r_g}$, the ISCO for a Schwarzschild BH. We use $1000$ grid cells, giving a grid spacing of ${\sim0.1}$. This grid spacing is sufficiently small that it is smaller than the $\beta$-grid spacing for all of our runs, thus ensuring that the results are resolution independent.

The time-step for the simulation was calculated on the fly. 
We require
\begin{equation}
    \label{eq:CFL}
    \Delta t < N_\text{CFL}(\Delta x)^2 \frac{4x}{3\alpha\Hcal^2} \, ,
\end{equation}
where $N_\text{CFL}$ is an {appropriate Courant} number. Due to the additional variation in $\alpha$, this is not a strict stability criterion. However, since the variation in $\alpha$ occurs on longer length scales that the diffusion grid, we do not expect this to be an issue. In our simulations we use ${N_\text{CFL}=0.25}$, a factor of 2 smaller than what would be required were this to be a true CFL number. Our simulations reveal no evidence of numerical instabilities which would be caused by having a too large time-step. In addition to the stability restriction, we further required that the time-step be sufficiently small so that the value of $\beta$ should not change by more than $1\%$ on average in any given time-step.

\begin{figure*}
	\centering
	\includegraphics[scale=0.59]{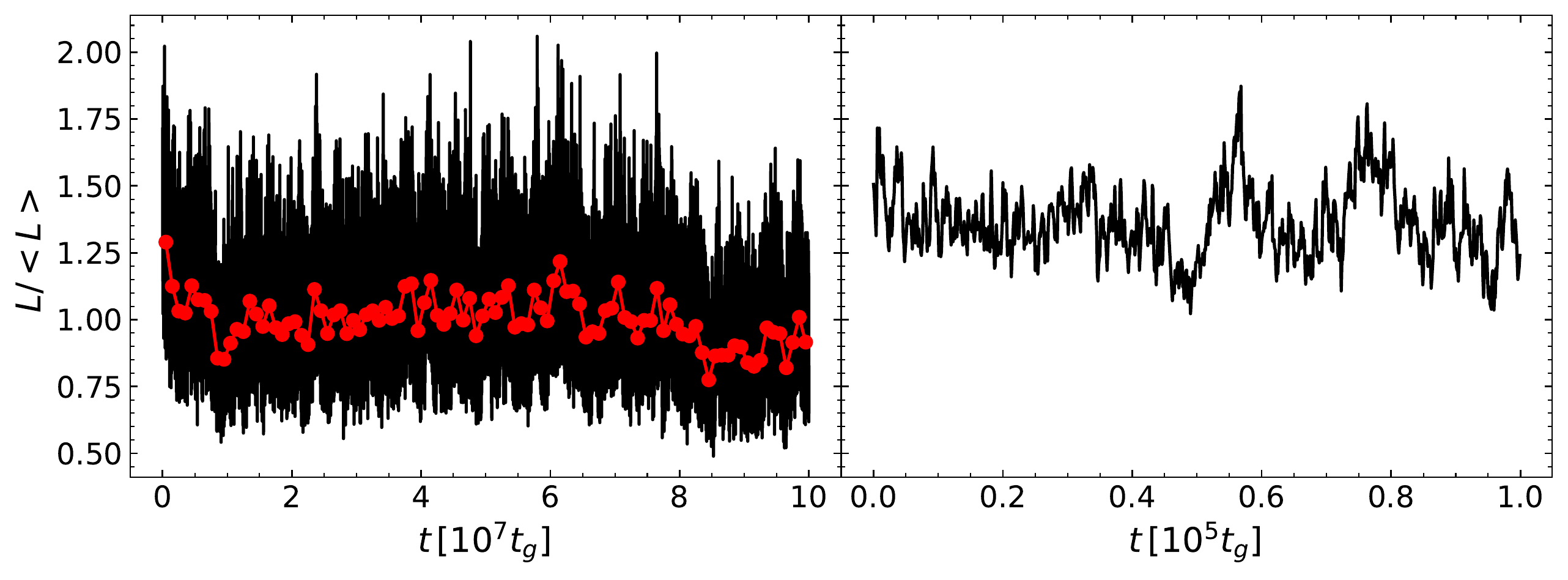}
	\caption[Synthetic light curve of the bolometric luminosity.]{The synthetic bolometric luminosity curve from our fiducial model. The luminosity is normalised to its mean value over the entire simulation. The left hand panel shows the entire $10^8t_g$ light curve in black, with data points every $100t_g$. The red points show the data binned into bins covering a time of $10^6t_g$ and so contain $10^4$ data points. The right hand panel shows the first $10^5t_g$ of the simulation on the same vertical scale, revealing the high frequency variation. These plots show that variation occurs over a large frequency range.}
	\label{fig:Lcurve}
\end{figure*}

{As boundary conditions, we force $\Psi$ to take its steady state value at the inner and outer edges of the disc. At the inner edge this means $\Psi=0$ for all time, which enforces a zero-torque boundary condition.}In order to avoid edge effects at the outer edge of the disc, the $50$ outermost cells are forced to have $\beta=0$. This ensures that there is a buffer zone which avoids unphysical effects from the interaction between the disc and the boundary condition. In addition, the {values of $\beta$ in the} other grid cells are multiplied by $\tanh(x^\dagger-x)$ where $x^\dagger$ is the position of the $50^\text{th}$ grid cell from the outer edge. This creates a smooth transition between the buffer zone and the main part of the disc.

All the simulations presented here have $\alpha_0=0.1$. {To organize the parameter exploration, we define a fiducial run that uses the exponential model for $\alpha(\beta)$ with ${\sqrt{\left<\beta^2\right>}=0.5}$, driving at the coherence length viscous timescale $t_{\nu,\text{c}}$ and a disc thickness of ${\Hcal=0.1}$.}

In our fiducial model, the driving timescale varies from ${\tilde{t}=10^{2.17}}$ at the inner edge of the disc to ${\tilde{t}=10^7}$ at the outer edge. Each run is set up with the steady state mass distribution (eq. \ref{eq:Psi_SS}) and $\beta=0$ throughout. While the steady state mass distribution depends on $\dot{M}$, it does so linearly. Since we only consider fractional changes in any the model results, the exact value used does not affect any of the results. The simulation is then {burnt}-in for ${\tilde{t}=10^7}$ to remove any transient response due to the initial conditions. The {production} run then extends {for a further} ${\tilde{t}=10^8}$, outputting the state of the disc at a cadence of ${\tilde{t}=10^2}$. This gives $10^6$ separate measurements of the state of the disc, covering the entire range of driving timescales present in the disc.

For simulations in which we vary the driving timescale, the burn-in time, run time and save cadence are adjusted proportionally, ensuring that the cadence is always less than the shortest driving timescale and the simulation runs for 10 times the longest driving timescale. 

\section{Fiducial Results} \label{sec:Fid_Res}

We start by discussing our fiducial model. Fig. \ref{fig:Lcurve} shows the bolometric light curve, calculated using eq. \eqref{eq:luminosity}. The plot reveals the expected fluctuating behaviour over a large frequency range. In this fiducial model, the driving timescale ranges from ${\tilde{t}=10^{2.17}}$ at the inner edge of the disc through to $10^7t_g$ at the outer edge. Looking at Fig. \ref{fig:Lcurve}, the fluctuations appear to cover this entire frequency range. Furthermore, the shape of the light curve appears to have sharp peaks and relatively flat troughs, hinting at log-normality.

\begin{figure*}
	\centering
	\includegraphics[scale=0.59]{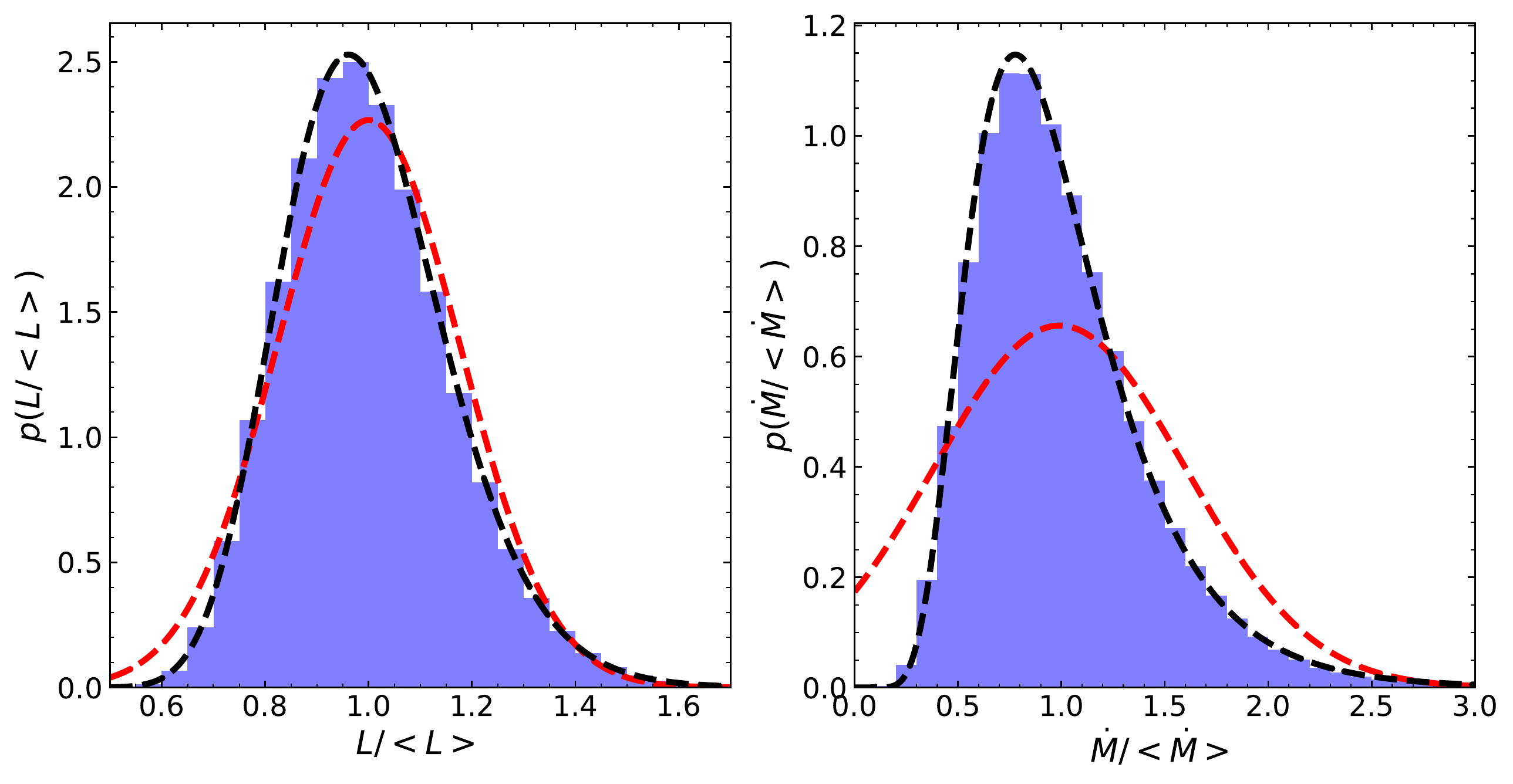}
	\caption{The shaded blue histograms show the distribution of the bolometric luminosity (left) and the mass accretion rate across the ISCO (right) for our fiducial model. In both cases, the quantities are normalised to their mean values over the entire simulation. The red and black dashed curves show the best fit normal and log-normal distributions respectively.}
	\label{fig:hist}
\end{figure*}

In order to test the log-normality or otherwise of the data, Fig. \ref{fig:hist} shows the probability density for the luminosity and also for the instantaneous mass accretion rate across the ISCO. We fit these data with normal and log-normal distributions
\begin{equation}
    \label{eq:normal_dist}
    f_\text{normal}(L;\mu,\sigma) = \frac{1}{\sqrt{2\pi}\sigma} \exp{\left[-\frac{(L-\mu)^2}{2\sigma^2}\right]} \, ,
\end{equation}
and
\begin{equation}
    \label{eq:lognormal_dist}
    f_\text{log-normal}(L;\mu,\sigma) = \frac{1}{\sqrt{2\pi}\sigma L} \exp{\left[-\frac{(\ln{L}-\mu)^2}{2\sigma^2}\right]} \, .
\end{equation}
The {fit to the PSD} was performed using the Markov Chain Monte Carlo (MCMC) package \textit{emcee} \citep{Foreman-Mackey+2013}, assuming that the data follow a Poisson distribution. Strictly, since each datum is not independent of the previous data points, this is not true, but given that the simulation is run for 10 times the longest driving timescale, this will be a very minor correction.

The results of the best fits of these distributions are also shown and Table \ref{tab:hist_values} shows the best fit parameters and the associated $\chi^2$ values for the four cases. The $\chi^2$ values show that the log-normal distribution gives a much better fit to the data for both $L$ and $\dot{M}$. It should be noted that the values of $\chi^2$ suggest that the actual distribution is not log-normal, despite the good visual fit. The reason for this is that the histograms contain a million individual measurements, and so even very small fractional differences between the synthetic observations and the model will give large values of $\chi^2$. Therefore, we are unable to conclude that our model gives a true log-normal output, but we can say that it appears to be a good approximation, and represents a vast improvement over a normal distribution.

Ordinarily, the MCMC fit produces error estimates on the best fit parameters. In this case of these fits, the $\chi^2$ values are very large and so even very small changes in the model parameters produce relatively large absolute changes in the value of $\chi^2$. The large values imply that the distributions are not accurate representations of the data but, in the case of the log-normal distribution, provide a good visual approximation. This extreme sensitivity to the changes in the model parameters mean that the errors produced by the MCMC fit are extremely small ($\sim10^{-4}$). Since these errors are not a true representation of the uncertainty of the fit they are not reported here.

\begin{table}
    \centering
    \caption{The best fit parameters for the plots in Fig. \ref{fig:hist} and the associated $\chi^2$ value divided by the degrees of freedom.}
    \label{tab:hist_values}
    \begin{tabular}{ccccc} \hline
        variable & model & $\mu$ & $\sigma$ & $\chi^2$/d.o.f. \\ \hline
        \multirow{2}{*}{$L$} & normal & $1.00$ & $0.176$ & $47400/29$ \\
        & log-normal & $-0.0132$ & $0.162$ & $301/29$ \\
        \multirow{2}{*}{$\dot{M}$} & normal & $0.991$ & $0.608$ & $363000/41$ \\
        & log-normal & $-0.0829$ & $0.411$ & $1930/41$ \\ \hline
    \end{tabular}
\end{table}

\begin{figure}
	\centering
	\includegraphics[width=\columnwidth]{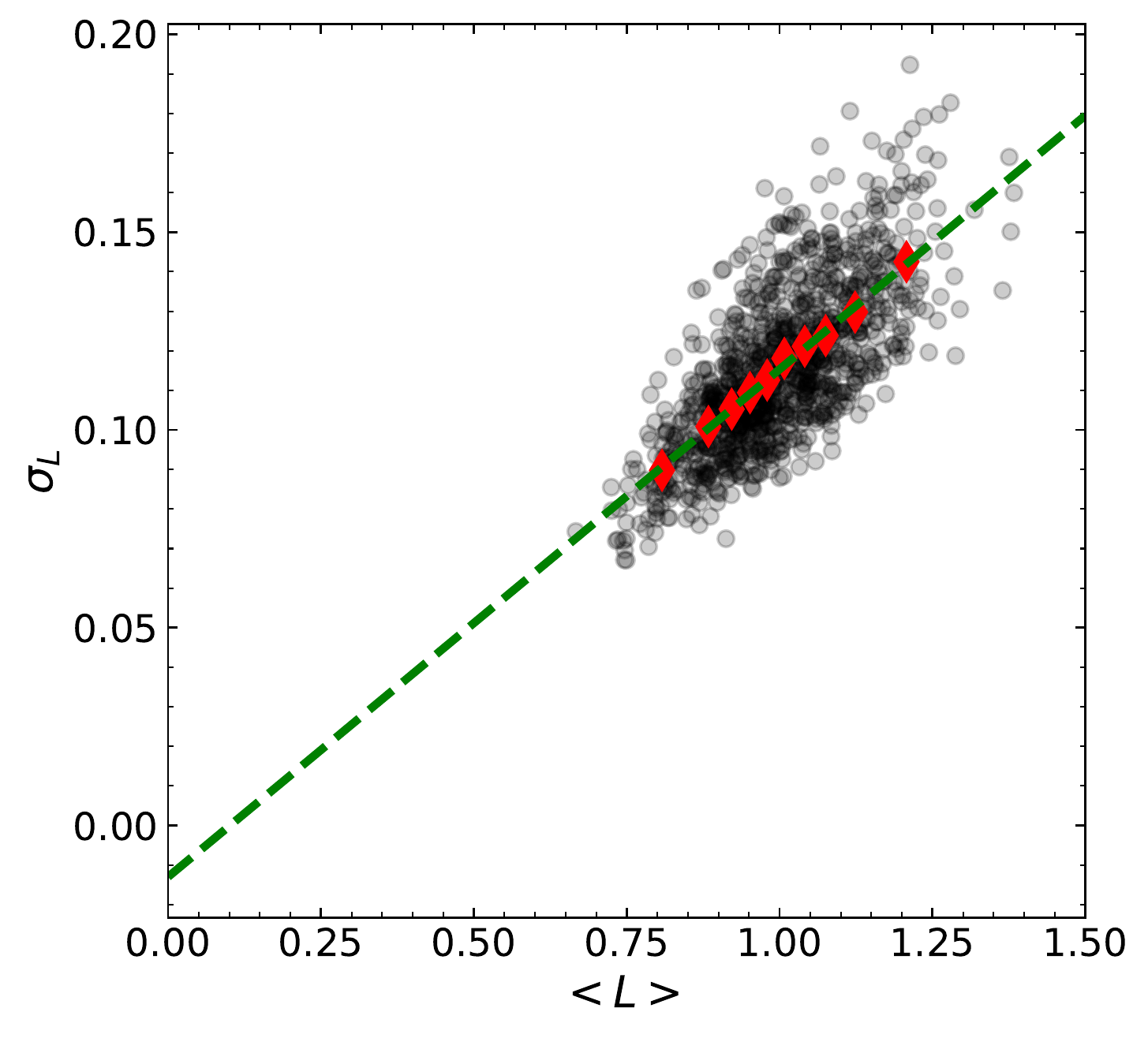}
	\caption{rms variation against the average luminosity of a section of the light curve. Each black point is calculated using $1000$ points ($10^5t_g$) of the light curve. The red points are the result of binning the black points into $10$ bins, each containing $100$ of the black points and are a visual guide only. The green line is the best fit straight line through the black points.}
	\label{fig:rms}
\end{figure}

Having established that the data are approximately log-normally distributed, we can now {examine} the rms-flux relationship. Given that these are equivalent \citep{Uttley+2005}, we expect a linear relationship. We divide the light curve into $1000$ bins, each containing $1000$ individual measurements covering a total of $10^5t_g$. In each of these bins, the average luminosity and rms variation is calculated. The results of this are shown in Fig. \ref{fig:rms}. To this data, we fit a straight line of the form
\begin{equation}
    \label{eq:rms_flux_line}
    \sigma_L = k(<L>-C) \, ,
\end{equation}
where $k$ and $C$ are constants, following the formalism of \citet{Uttley&McHardy2001}. The best fit line is also shown in Fig. \ref{fig:rms} with best fit parameters of ${k=0.128\pm0.004}$ and ${C=0.10\pm0.03}$.

We can extend our analysis by examining our results in Fourier space. {For this}, the total light curve is divided into ten sections, each of which covers $10^7t_g$ (the longest driving timescale of the disc) and contains $10^5$ individual data points. For each of the time series $s_i(t)$ we take the fast Fourier transform (FFT) $S_i(f)$ where the subscript $i$ represents the $i^\text{th}$ segment of the light curve. We can then calculate the power spectral density (PSD) by averaging over the ten segments as
\begin{equation}
    \label{eq:PSD}
    \text{PSD} = \langle|S_i(f)|^2\rangle = \langle S_i(f)S_i^*(f)\rangle \, .
\end{equation}
The PSD is well modelled by a broken power-law of the form
\begin{equation}
    \label{eq:broken_powerlaw}
    \text{PSD} \propto \begin{cases}
    f^{m_1}\, ,&f<f_\text{break} \\
    f^{m_2}\, ,&f>f_\text{break}
    \end{cases}
    \, ,
\end{equation}
where $m_1$ and $m_2$ are gradients in log space and $f_\text{break}$ is the break frequency between the two regimes. The broken power-law can be fit to the data using the same MCMC code as before. The parameters $m_1$, $m_2$ and $f_\text{break}$ were all fit, along with an additive constant (in log space) which is not physically relevant. It was also required that the model be continuous at $f_\text{break}$.

\begin{figure}
	\centering
	\includegraphics[width=\columnwidth]{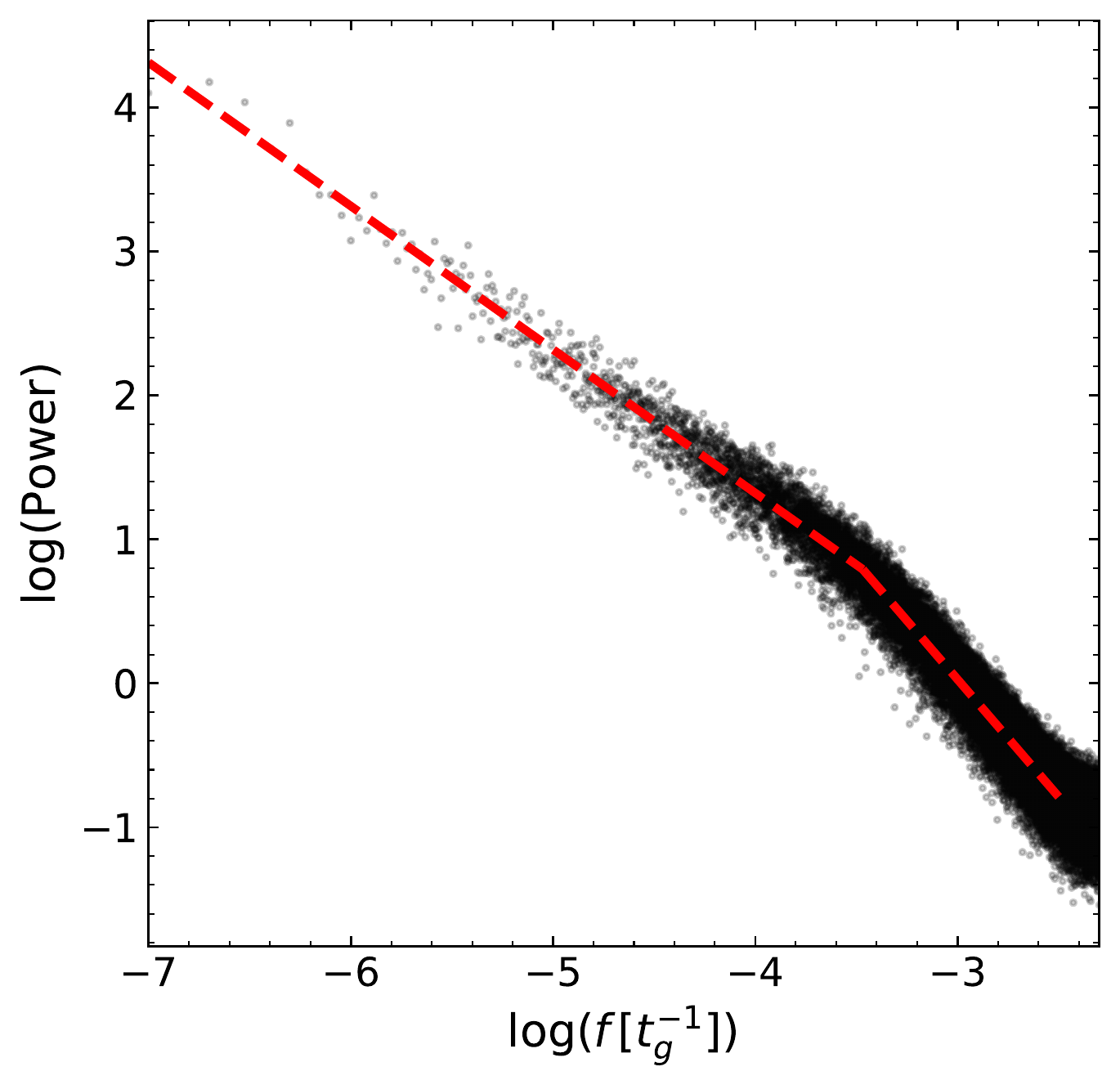}
	\caption{PSD for the bolometric luminosity from our fiducial model (black points). The dashed red line shows the best fit broken power-law. The domain of the broken power-law corresponds to the range of frequencies used in performing the fit to the data. This clearly shows the flattening at the high frequency end which is excluded from the fit.}
	\label{fig:PSD}
\end{figure}

The {luminosity} PSD and the best fit broken power-law are shown in Fig. \ref{fig:PSD}. This figure shows that, at the very highest frequencies, the power spectrum becomes flat. This is an artifact of the Fourier transform process and is not physical. For this reason, when fitting the broken power law to the data, we only consider frequencies {$<10^{-2.5}\,t_g^{-1}$}. The fit was obtained with $m_1=-0.996\pm0.016$, $m_2=-1.631\pm0.009$ and ${\log( f_\text{break}/t_g^{-1})=-3.47\pm0.02}$. The low frequency slope is consistent with $-1$ which represents flicker noise, as predicted by analytic models of $\alpha$ discs \citep{Lyubarskii1997}. A gradient of $-1$ means that there is constant $fP(f)$ power in all frequencies. The low frequency power must be created in the outer regions of the disc. Since the luminosity is predominantly produced in the inner regions, this power must be transferred inwards through the propagating fluctuations. For frequencies greater than the break frequency, there is a decreasing amount of power at higher frequencies. This is a result of the fact that there are no fluctuations being created at these frequencies and so the power drops rapidly.

\begin{figure}
	\centering
	\includegraphics[width=\columnwidth]{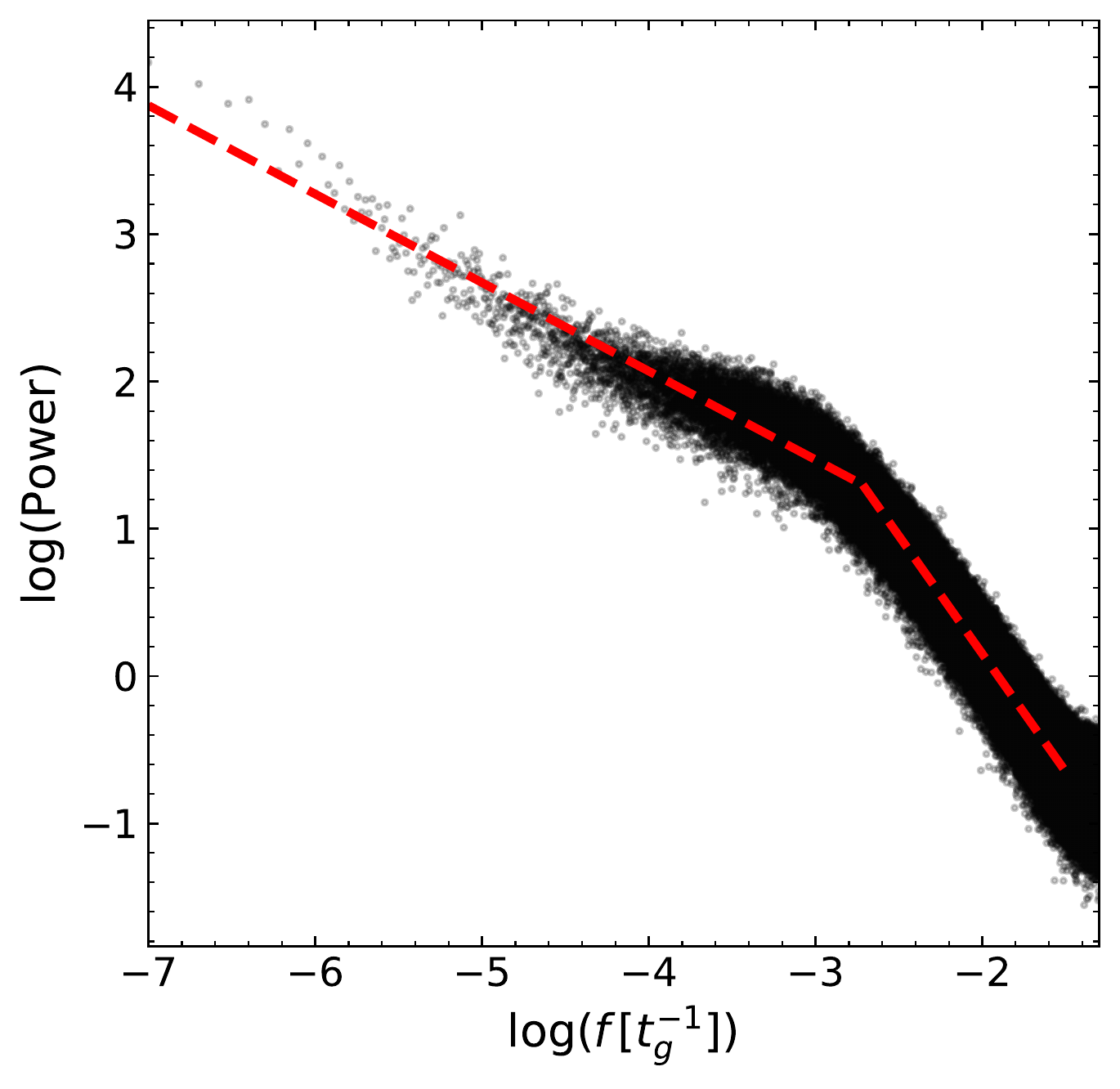}
	\caption{As in Fig. \ref{fig:PSD} but for the accretion rate across the ISCO.}
	\label{fig:PSD_Mdot}
\end{figure}

We can also calculate the PSD for the instantaneous accretion rate across the ISCO. The break frequency is larger for the accretion rate than the luminosity. It is therefore necessary to consider a higher time resolution in order to reveal the full shape of the PSD. To do this, we have performed a high resolution run of our fiducial model. In this run, the output cadence is $\tilde{t}=10^1$ as compared to the standard $10^2$ { but only the bolometric luminosity and the accretion rate across the ISCO are saved as opposed to the entire state of the disc in the lower resolution runs.} The PSD is shown in Fig. \ref{fig:PSD_Mdot}. We have fit the $\dot{M}$ PSD with a broken power-law in the same way as the luminosity. The best fit was obtained with $m_1=-0.600\pm0.007$, $m_2=-1.600\pm0.002$ and ${\log( f_\text{break}/t_g^{-1})=-2.718\pm0.006}$.

A comparison of Figs. \ref{fig:PSD} and \ref{fig:PSD_Mdot} reveals two interesting results. Firstly, the accretion rate PSD appears visually to have three rather than two sections, with a  flatter domain in { intermediate frequencies}. This is reflected in the smaller absolute value of $m_1$, representing a shallower low frequency slope. The possible reasons behind this as explored in $\S$\ref{sec:timescale}.

Secondly, the break frequency for the luminosity PSD is smaller by a factor of ${\sim5}$ than that for the accretion rate. When considering the accretion rate across the ISCO, fluctuations created at all radii will propagate inwards and effect the variability. In general, fluctuations propagate inwards and not outwards. If we consider the dissipation at ${x=2x_*}$, it will be affected by fluctuations produced at ${x>2x_*}$ but not those from ${x<2x_*}$. This means that the PSD of the dissipation at larger radii will have lower break frequencies than that at smaller radii. Since the luminosity is produced over a small but extended region, it will consist of weighted combinations from a number of radii, all of which would be expected to have lower break frequencies than that accretion rate at the ISCO. It is therefore unsurprising that this difference between the break frequencies is seen here. It is worth noting that \citet{Mushtukov+2018} considered an analytic model with outwardly as well as inwardly propagating fluctuations. While this does have an effect, the inwards propagation is still the dominant effect.

{Given that we have argued that the accretion rate PSD is sensitive to fluctuations in the whole disc, we might expect that the break frequency be set by the largest driving frequency, found at the inner edge of the disc. Instead we see that the break frequency is ${\log( f_\text{break}/t_g^{-1})=-2.718\pm0.006}$ while the largest driving frequency is ${\log( f_\text{drive,min}/t_g^{-1})=-2.167}$. A similar result was found by \citet{Ingram&Done2011} but its explanation first requires a consideration of coherence within the disc.}

Within the propagating fluctuations model, it is assumed that two regions of the disc can only communicate with each other at frequencies below that set by the viscous diffusion time between them. Any higher frequency signal is smoothed out by the diffusion process and so not passed on to smaller radii. One natural way to examine this is to look at the coherence in the dissipation at different radii. Following \citet{Nowak+1999}, two time series $s(t)$ and $h(t)$ can be related by a transfer function $t_r(\tau)$ through
\begin{equation}
    \label{eq:transfer_function}
    h(t) = \int_{-\infty}^\infty t_r(t-\tau)s(\tau)\text{d}\tau \, .
\end{equation}
This is a convolution and so becomes a multiplication in Fourier space
\begin{equation}
    \label{eq:transfer_function_Fourier}
    H(f) = T_r(f)S(f) \, .
\end{equation}
If this transfer function is specified then, given either $s(t)$ or $S(f)$, we can calculate $h(t)$ or $H(f)$. The coherence function measures how much $T_r(f)$ varies between measurements. It is defined as
\begin{equation}
    \label{eq:coherence_function}
    \gamma^2(f) = \frac{|\langle S^*(f)H(f)\rangle|^2}
    {\langle|S(f)|^2\rangle\langle|H(f)|^2\rangle} \, ,
\end{equation}
where the averaging occurs over the ten segments of the light curve, as in eq. \eqref{eq:PSD}, and we have dropped the subscript $i$ for {clarity}. A value of $\gamma^2=1$ shows that the two time series are completely coherent and $\gamma^2=0$ shows complete incoherence.

\begin{figure*}
	\centering
	\includegraphics[scale=0.59]{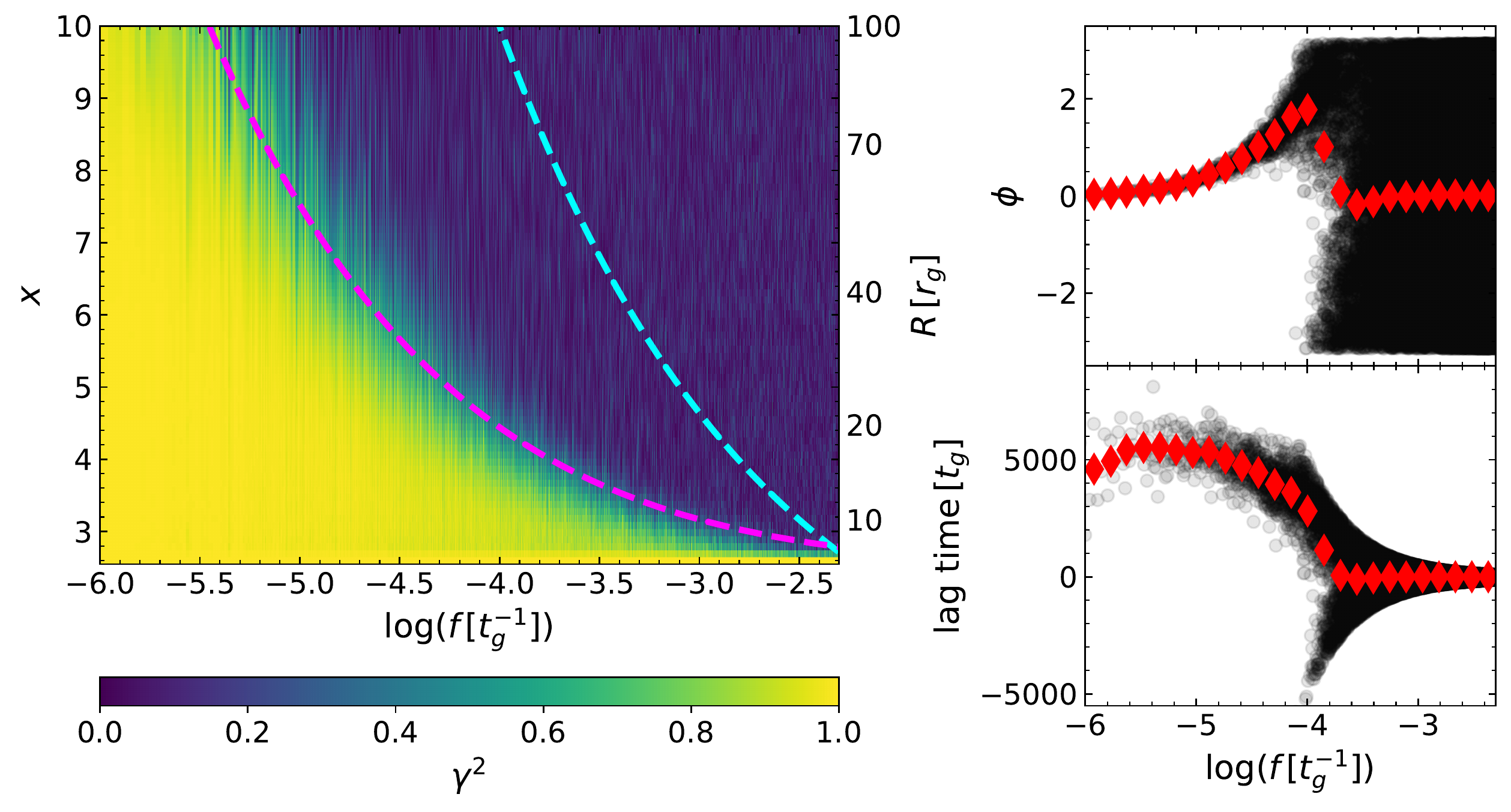}
	\caption{\textit{Left:} Coherence map for the local dissipation rate, relative to the inner edge of the disc. Also shown are the local driving frequency (cyan) and the viscous diffusion time to reach the inner edge in steady state (magenta). The figure clearly shows the change from coherence at low frequencies to incoherence at high frequencies. \textit{Right:} The phase (top) and time (bottom) lag between the local dissipation at $x=5$ and $x=\sqrt{6}$. The black circles show all the unbinned data while the red diamonds show the data having been binned. Positive lags correspond to the dissipation at $x=\sqrt{6}$ lagging behind that at $x=5$. The high frequency envelope in the lag time is caused by the fact that $\phi\in[-\pi,\pi]$. Therefore, the lag time must be bound by ${\pm1/2f}$.}
	\label{fig:coherence_map}
\end{figure*}

We take $h(t)$ to be the dissipation rate in the second grid cell at the inner edge of the disc. The reason for not using the first grid cell is that it has $\Sigma=0$ from the zero-torque boundary condition and so never has any dissipation. Taking $s(t)$ as the dissipation rate in another radius, we can calculate the coherence function between any radius at the inner edge of the disc. This gives $\gamma^2(x,f)$ which is shown in the left panel of Fig. \ref{fig:coherence_map}.

We can calculate the viscous diffusion time in steady state and use this to test the assumption that coherence appears at frequencies below this. We can combine eqs. \eqref{eq:thin_radial_velocity} and \eqref{eq:Sigma_SS} to find the radial velocity is steady state
\begin{equation}
    \label{eq:SS_radial_velocity}
    u_R = \dev{R}{t} = \frac{-3\nu}{2\left(R-\sqrt{RR_*}\right)} \, ,
\end{equation}
where ${R_*=6r_g}$ is the inner edge of the disc. We can then use eq. \eqref{eq:alpha_viscosity_x} and convert this into $x$ and $\tilde{t}$, giving
\begin{equation}
    \label{eq:SS_radial_velocity_x}
    \dev{x}{\tilde{t}} = \frac{-3\alpha\Hcal^2}{4\left(x^2-xx_*\right)} \, .
\end{equation}
We can then simply integrate eq. \eqref{eq:SS_radial_velocity_x} from $x$ to $x_*=\sqrt{6}$ to give the viscous travel time
\begin{equation}
    \begin{aligned}
        \label{eq:SS_travel_time}
        \tilde{t}_\text{travel} &= \frac{4}{3\alpha\Hcal^2}
        \left(\frac{1}{3}x^3 - \frac{1}{2}x^2x_* + \frac{1}{6}x_*^3\right) \\
        &= \frac{4}{3\alpha\Hcal^2}
        \left(\frac{1}{3}x^3 - \frac{\sqrt{6}}{2}x^2 + \sqrt{6}\right) \, .
    \end{aligned}
\end{equation}

{The frequency corresponding to this time} is shown in Fig. \ref{fig:coherence_map} in magenta, and follows the divide between coherence and incoherence very closely, supporting the assumption that high frequency fluctuations (relative to the viscous travel time) are not passed through the disc.

For regions of Fig. \ref{fig:coherence_map} which are almost perfectly coherent, this does not mean that the two time series $h(t)$ and $s(t)$ are the same, but that they are related by an (approximately) constant transfer function. Therefore, we can write ${H(f)=S(f)A(f)e^{-i\phi(f)}}$, where $A(f)$ is a real, frequency dependent constant and $\phi(f)$ is the phase by which $h(t)$ lags $s(t)$. This phase corresponds to a time lag of ${t_\text{lag}(f)=\phi(f)/2\pi f}$.

The {temporal-frequency dependent} phase and time lags between $x=5$ and $x=\sqrt{6}$ are shown in the right panel of Fig. \ref{fig:coherence_map}. At low frequencies there is very little scatter in the phase, consistent with the dissipation being coherent. At higher frequencies, the scatter increases and coherence is lost. The calculation of the phase difference can only produce values that lie between ${-\pi}$ and $\pi$. At high frequencies, the phase appears to be entirely random, with the binned averages of the large number of random values lying close to $0$ as expected. The cross over between coherence and incoherence occurs just below $10^{-4}/t_g$, completely consistent with the coherence map at $x=5$ shown in Fig. \ref{fig:coherence_map}.

The time lag appears to be approximately constant at low frequencies and equal to ${\sim5000 t_g}$. The viscous travel time from $x=5$ to $x=\sqrt{6}$ (eq. \ref{eq:SS_travel_time}) is ${\sim18000 t_g}$, more than three times larger than the lag time. Intuitively, if the lag is caused by propagating fluctuations, we might expect these times to be similar, especially since the viscous travel time fits well with the coherence map in Fig. \ref{fig:coherence_map}. A similar result was found by \citet{Cowperthwaite&Reynolds2014}.

\begin{figure*}
	\centering
	\includegraphics[scale=0.59]{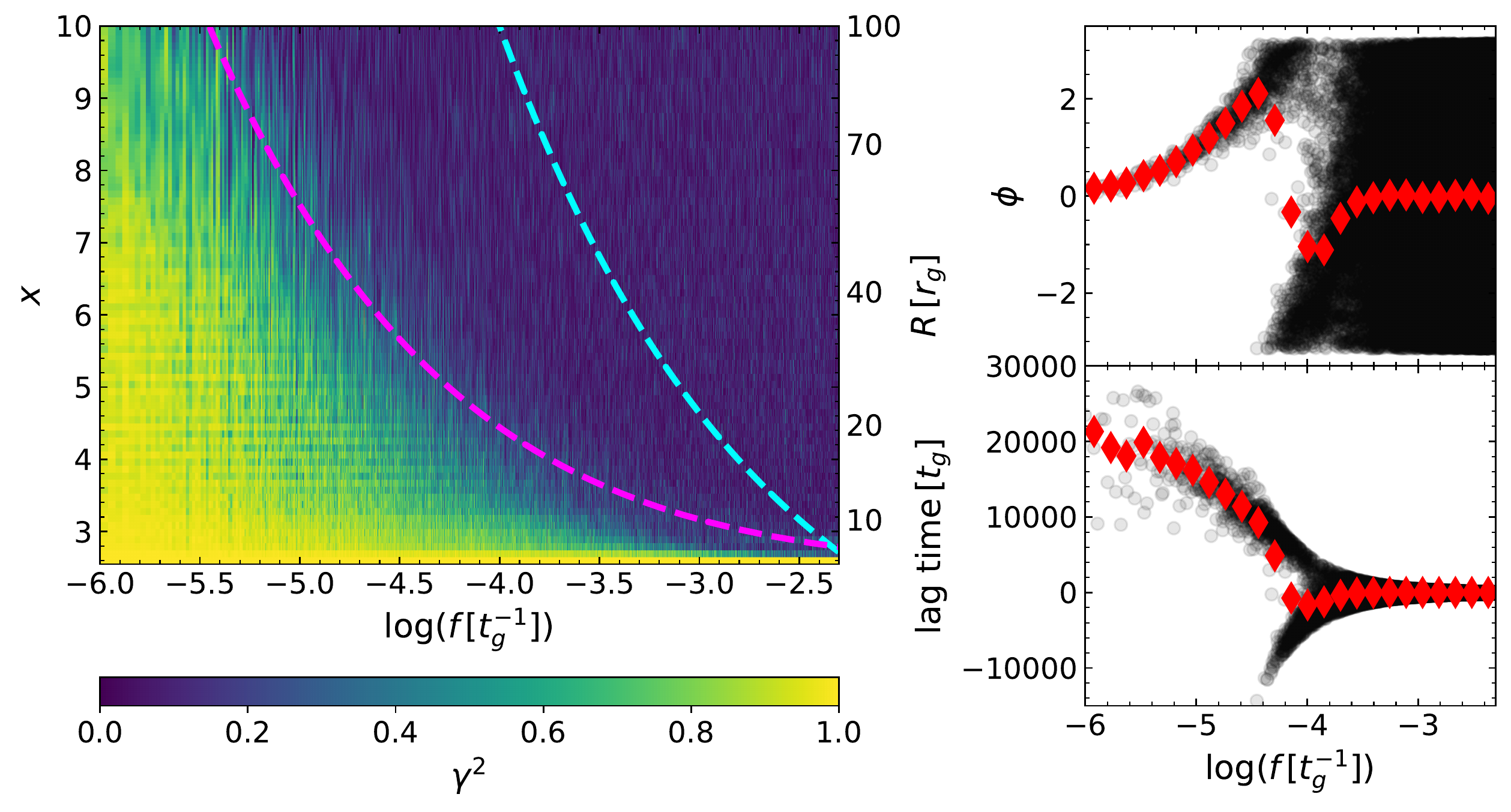}
	\caption{As for Fig. \ref{fig:coherence_map} but for the local accretion rate rather than the dissipation.}
	\label{fig:coherence_map_Mdot}
\end{figure*}

We can also look at the coherence and lag of the local accretion rate. These results are shown in Fig. \ref{fig:coherence_map_Mdot}. There are two key differences between the accretion rate and dissipation. Firstly, the accretion rate becomes coherent at lower frequencies than the dissipation. The viscous travel time still provides a reasonable guide to the crossover between coherence and incoherence, but this crossover appears to be shifted to slightly lower frequencies. The second difference is that the lag time between $x=5$ and $x=\sqrt{6}$ is now $\sim20000t_g$. This is very similar to the travel time between the two radii.

{We are now able to return to the discussion of Fig. \ref{fig:PSD_Mdot} and the discrepancy between the break frequency of the accretion rate PSD and the largest driving frequency. In their model \citet{Ingram&Done2011} found that, since fluctuations on timescales shorter than the viscous travel time are incoherent, coherence between the variability on the fastest timescales was reduced, suppressing the high frequency power. This resulted in a lower break frequency than would be naively expected from simply considering the driving frequency at the inner edge of the disc. In our fiducial simulation, the driving timescale is shorter than the viscous travel time in almost the entire disc (see Fig. \ref{fig:coherence_map}). It is therefore unsurprising that some of the highest frequency power would be lost and this explanation is entirely consistent with what we have found in our simulations.}

This difference between the lag in the accretion rate and the dissipation is worthy of further discussion. The similarity between the viscous travel time and the lag time in the accretion rate implies that the accretion rate fluctuations propagate through the disc viscously, as would be expected. In the case of the dissipation, there must be an additional process by which information can be transferred more quickly. The most obvious mechanism by which this could happen is via viscous torques. In steady state, the local dissipation rate is not equal to the rate at which gravitational potential energy is liberated. Instead, some of the liberated energy is transferred outwards before it is dissipated at larger radii than it is dissipated. In the outer regions of a steady state disc, only a third of the dissipated energy comes from local liberation of gravitational potential energy, the remainder comes via viscous torques from the inner disc. This process of the transfer of energy could ensure that information can be transferred through the disc at a faster rate than viscous travel, thus creating the shorter lag time in the local dissipation between radii.

It is worth noting that, as the accretion rate becomes incoherent at high frequencies, the phase and the lag time briefly become negative. This means that the accretion rate at $x=5$ is lagging behind that at $x=\sqrt{6}$. The work of \citet{Mushtukov+2018} added outwardly propagating fluctuations to the standard propagating fluctuations model. They predicted that this could create negative time lags between radii at high frequencies, similar to what we have observed here.

\section{Effect of Model Parameters}

While the results presented in $\S$\ref{sec:Fid_Res} are consistent with our expectations of the model, there is nothing special about the input parameters chosen for the fiducial model. In this section, we therefore explore the effect of varying several of these parameters on the simulation results.  

\subsection{Magnitude and distribution of $\alpha$-fluctuations}

\begin{table*}
    \centering
    \caption{Parameters values for the best fit normal and log-normal distributions to the probability distribution of the bolometric luminosity for four different models for $\alpha(\beta)$. Also shown are the parameters of the best fit broken power-laws for their respective PSDs.}
    \label{tab:hist_values_beta}
    \begin{tabular}{ccccccccccc} \hline
        \multirow{2}{*}{$\beta$ model} & \multirow{2}{*}{$\sqrt{\left<\beta^2\right>}$} & \multicolumn{3}{c}{normal} & \multicolumn{3}{c}{log-normal} & \multicolumn{3}{c}{PSD} \\
        & & $\mu$ & $\sigma$ & $\chi^2$/d.o.f. & $\mu$ & $\sigma$ & $\chi^2$/d.o.f. & $m_1$ & $m_2$ & $\log(f_\text{break}/t_g^{-1})$ \\ \hline
        \multirow{2}{*}{linear} & $0.2$ & $1.00$ & $0.0703$ & $6130/23$ & $-0.002$ & $0.0698$ & $517/23$ & $-1.000\pm0.015$ & $-1.682\pm0.009$ & $-3.464\pm0.019$ \\
        & $0.5$ &$0.999$ & $0.291$ & $101000/24$ & $-0.0309$ & $0.249$ & $671/24$ & $-0.975\pm0.015$ & $-1.748\pm0.011$ & $-3.357\pm0.018$ \\
        \multirow{2}{*}{exponential} & $0.2$ & $1.00$ & $0.0688$ & $5816/23$ & $-0.00232$ & $0.0682$ & $99.2/23$ & $-1.014\pm0.014$ & $-1.668\pm0.009$ & $-3.47\pm0.02$ \\
        & $0.5$ & $1.00$ & $0.176$ & $47400/29$ & $-0.0132$ & $0.162$ & $301/29$ & $-0.996\pm0.016$ & $-1.631\pm0.009$ & $-3.47\pm0.02$ \\ \hline
    \end{tabular}
\end{table*}

We first consider the effect of varying the value of $\sqrt{\left\langle\beta^2\right\rangle}$ and the choice between the linear and exponential model for $\alpha(\beta)$. We use four simulations to do this, with ${\sqrt{\left\langle\beta^2\right\rangle}=\{0.2,0.5\}}$ for both the linear and exponential model. Note the exponential model with ${\sqrt{\left\langle\beta^2\right\rangle}=0.5}$ is our fiducial model.

When considering the linear models, it is important to quantify the effect of the floor applied to the value of $\beta$ to ensure that $\alpha$ is always positive. Eq. \eqref{eq:coherence_recurrence} can be inverted to show that there are ${\sim77}$ independent $\beta$ cells in these simulations (which have ${\Hcal=0.1}$. The floor is applied whenever ${\beta<-1}$. If ${\sqrt{\left\langle\beta^2\right\rangle}=0.2}$ then this is a one-sided $5\sigma$ value and occurs with a probability of ${2.87\times10^{-7}}$. With $77$ independent $\beta$ values and a total simulation time of ${10^8\,t_g}$, this means that we expect the floor to be in use somewhere in the domain for a total time of ${2210\,t_g}$ or ${0.002\%}$ of the total time. We can therefore say, to a very good approximation, that the floor will have a minimal effect on the simulation with ${\sqrt{\left\langle\beta^2\right\rangle}=0.2}$.

We can contrast this with the case when ${\sqrt{\left\langle\beta^2\right\rangle}=0.5}$ which only requires a $2\sigma$ value for the floor to be used. This occurs with a probability of $0.0227$. In this case, the probability that the floor is in use in at least one of the $\beta$ cells at any given time is $0.830$. Therefore this simulation will be influenced strongly by the floor.

Table \ref{tab:hist_values_beta} shows the best fit parameters and reduced $\chi^2$ values produced by fitting the luminosity distribution with both a normal and log-normal distribution. It also shows the best fit parameters for the broken power-law fit to the PSD. From these data we can identify a few interesting results. Firstly, we can see that those simulations with larger values of $\sqrt{\left\langle\beta^2\right\rangle}$ produce more variability as can be seen by their larger best fit values of $\sigma$. This is not surprising as larger fluctuations in the viscosity would be expected to produce more variability. Comparing the linear and exponential models with ${\sqrt{\left\langle\beta^2\right\rangle}=0.2}$, we can see that their results are very similar. This is not unexpected since, for small values of $\beta$, ${e^\beta\approx1+\beta}$. 
Generally, the exponential model is better fit by a log-normal distribution than the linear model. However, in all cases the log-normal provides a much better fit that the normal distribution. It is interesting to note that the linear model with ${\sqrt{\left\langle\beta^2\right\rangle}=0.2}$ favours the log-normal fit. For this model, we have argued that the $\beta$ floor will have only a very small effect and so the value of $\alpha$ used in the simulation should be normally distributed around $\alpha=\alpha_0$. Therefore, the log-normality is a direct result of the evolution of the diffusion equation and does not appear to be imposed by the choice of distribution for the viscosity.

The parameters for the broken power-law show broad consistency between the models, with the exception of the linear model with ${\sqrt{\left\langle\beta^2\right\rangle}=0.5}$. This is the model in which the $\beta$ floor has a near-constant impact on the simulation. The effect this has on the value of the break frequency is especially noteworthy and suggests that the floor could have a large impact on the values of observable values.

\subsection{Disc Thickness}

The fiducial model used an aspect ratio of ${\Hcal=0.1}$. This is larger than the value expected of {many} real thin discs (e.g. \citealt{Frank+2002}). However, reducing the aspect ratio of the disc significantly increases computational expense. Since the coherence length of $\beta$ is proportional to $\Hcal$ (eq. \ref{eq:coherence_length}), reducing $\Hcal$ increases the required number of grid cells to resolve the viscosity adequately. This in turn drastically reduces the maximum time-step allowed. {However, our goal here is to gain insight into the behaviour of these models rather than describe real accretion discs, so it is unnecessary to explore thinner systems.} Instead, we here explore models with ${\Hcal\in[0.1,0.5]}$. These thicker discs would, in reality, not behave as true thin discs {since the radial advection of thermal energy would become important}, but exploring this parameter space allows us to draw general conclusions about the effect of disc thickness in our simple models.

\begin{figure*}
	\centering
	\includegraphics[scale=0.59]{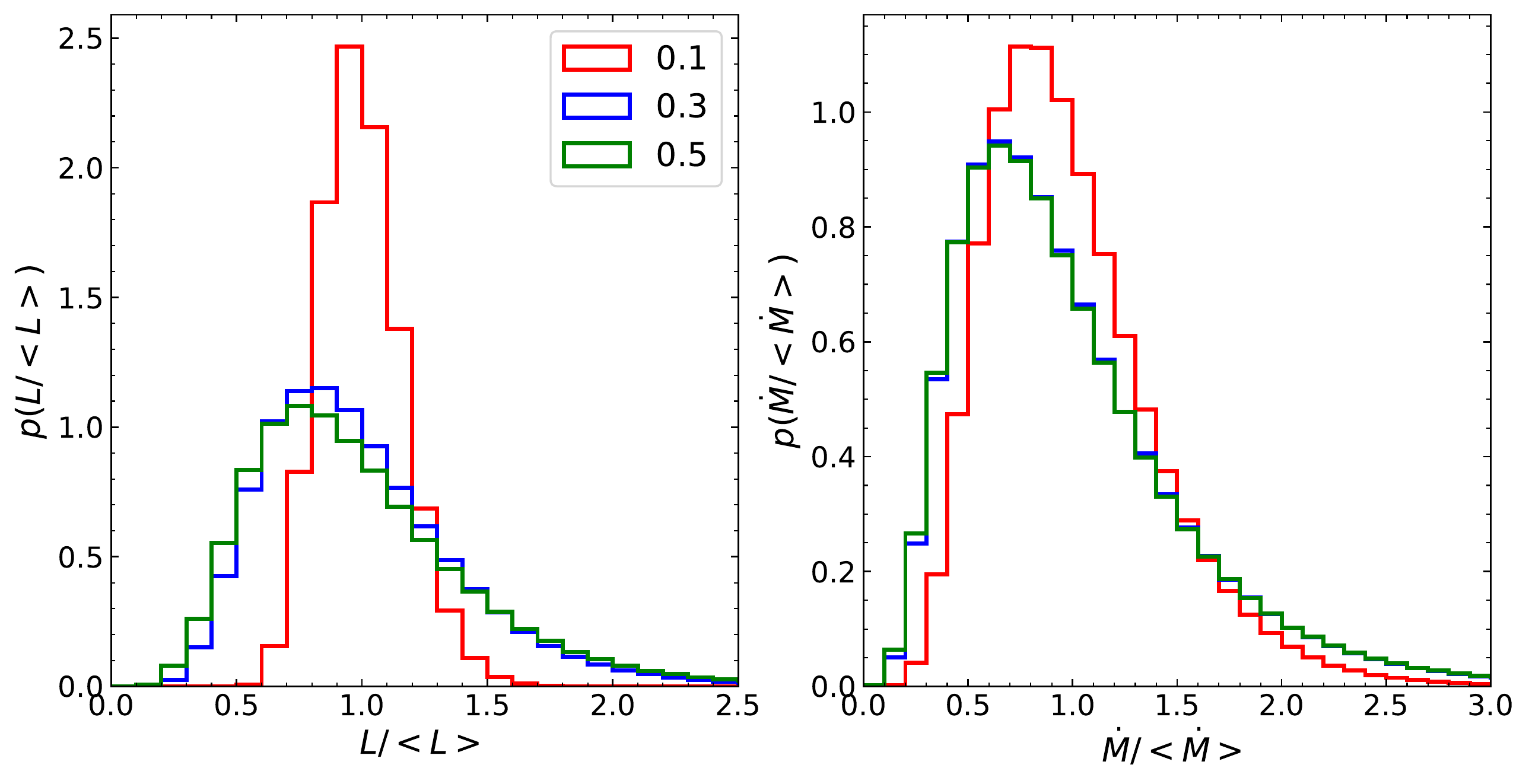}
	\caption{Probably distributions for the bolometric luminosity (left) and accretion rate across the ISCO (right) for three different disc aspect ratios.}
	\label{fig:aspect_hist}
\end{figure*}

Fig. \ref{fig:aspect_hist} shows the distribution of bolometric luminosity and instantaneous accretion rate across the ISCO for different values of $\Hcal$. It clearly shows that increasing $\Hcal$ greatly increases the width of the luminosity distribution. While the distribution of $\dot{M}$ also increases, it does so to a much smaller degree. To understand this effect, we need to consider that, while the accretion rate depends on the behaviour at the very inner edge of the disc, the luminosity comes from an extended area of the disc. By substitution of the steady state mass distribution in eq. \eqref{eq:Psi_SS} into the luminosity integral in eq. \eqref{eq:luminosity}, we can show that the half-light radius is $x_\text{hl}=2x_*$ (i.e. the radius at which half the total luminosity comes from $x<x_\text{hl}$) which corresponds to ${R=4R_*}$.

Inverting eq. \eqref{eq:coherence_recurrence} gives
\begin{equation}
    \label{eq:number_cells}
    n = \frac{\log(x_\text{hl}/x_*)}{\log\left(1+\frac{1}{2}\Hcal\right)} \, ,
\end{equation}
for the number of viscous coherence lengths between the inner edge of the disc and the half-light radius. Calculating this value for $\Hcal=0.1$, $0.3$ and $0.5$ gives $n=14.2$, $4.96$ and $3.11$ respectively. Very crudely, each coherent cell can be thought of as fluctuating independently, with the total luminosity being given by their sum. Therefore, when there are more cells there will be a greater cancelling of the fluctuations, producing a narrower distribution of luminosity as seen in Fig. \ref{fig:aspect_hist}. This argument does not apply to the accretion rate across the ISCO as there is no combination of different cells involved. Therefore, we can intuitively understand why the accretion rate is affected to a much smaller extent compared to the luminosity. For discs {with thicknesses appropriate for many real CV and AGN}, $\Hcal\sim0.01$ \citep{Frank+2002}, there are $\sim139$ coherent cells within the half-light radius. We would therefore expect the thermal disc luminosity to be considerably less variable, assuming the trend observed can be extrapolated.

This extra variability in thicker discs is mirrored in the rms-flux relation. The general shape seen in Fig. \ref{fig:rms} is independent of $\Hcal$, but the slope increases for larger $\Hcal$. This is exactly as expected given that larger slopes imply greater variability as seen in Fig. \ref{fig:aspect_hist}. For $\Hcal=0.3$, the slope of the rms-flux relation is ${0.270\pm0.005}$ compared with ${0.128\pm0.004}$ for $\Hcal=0.1$. This result is not a surprise given the close relation between the log-normality and the rms-flux relation.

The {luminosity and accretion rate} PSDs of these simulations are also well modelled by a broken power-law and the corresponding best fit parameters are given in Table \ref{tab:aspect_PSD}. { In order to calculate the accretion rate PSDs, it was again necessary to run simulations with the higher output cadence of ${\tilde{t}=10t_g}$, saving only summary information rather than the entire state of the disc.}

While the low frequency slope of the luminosity PSD in our fiducial model is consistent with $-1$, the disc models with greater values of $\Hcal$ show statistically significant disagreements with a $-1$ slope. However, in absolute terms the difference is small and, since our model contains non-linearity which is not present under the assumptions made by \citet{Lyubarskii1997}, not overly significant.

The {luminosity} PSDs reveal an interesting trend. The {results} in Table \ref{tab:aspect_PSD} clearly show that larger values of $\Hcal$ results in smaller values of $f_\text{break}$. Since the driving frequency is not changing between the models, this can not be the cause. As we have seen, thicker discs show more variability than thinner discs. This extra variability in thicker discs can increase the size of the region of the disc from which the majority of the luminosity comes from. This increase in the radius of the emitting region means that the high frequency fluctuations (from the inner disc) have less of an impact on the luminosity and so the break frequency is lower than in thinner discs. To a first approximation, we find a linear fitting formula between the disc thickness and the break frequency of
\begin{equation}
    \label{eq:aspect_fit}
    \log(f_\text{break}/t_g^{-1}) = m\Hcal + c\, ,
\end{equation}
with ${m=-0.31\pm0.09}$ and ${c=-3.47\pm0.02}$.

\begin{table*}
    \centering
    \caption{Parameters for the best fit broken power-laws for the { luminosity and accretion rate PSDs} given by simulations with different values of $\Hcal$.}
    \label{tab:aspect_PSD}
    \begin{tabular}{cccc>{}c>{}c>{}c} \hline
        \multirow{2}{*}{$\Hcal$} & \multicolumn{3}{c}{$L$} & \multicolumn{3}{c}{$\dot{M}$} \\
        & $m_1$ & $m_2$ & $\log(f_\text{break}/t_g^{-1})$ & $m_1$ & $m_2$ & $\log(f_\text{break}/t_g^{-1})$ \\ \hline
        $0.1$ & $-0.996\pm0.016$ & $-1.631\pm0.009$ & $-3.47\pm0.02$ & $-0.600\pm0.007$ & $-1.600\pm0.002$ & $-2.718\pm0.006$ \\
        $0.2$ & $-1.102\pm0.018$ & $-1.684\pm0.010$ & $-3.57\pm0.02$ & $-0.903\pm0.006$ & $-1.664\pm0.003$ & $-2.507\pm0.007$ \\
        $0.3$ & $-1.16\pm0.02$ & $-1.677\pm0.010$ & $-3.54\pm0.04$ & $-1.000\pm0.006$ & $-1.672\pm0.003$ & $-2.455\pm0.009$ \\
        $0.4$ & $-1.146\pm0.018$ & $-1.623\pm0.008$ & $-3.59\pm0.03$ & $-1.042\pm0.006$ & $-1.665\pm0.003$ & $-2.453\pm0.009$ \\
        $0.5$ & $-1.15\pm0.02$ & $-1.587\pm0.009$ & $-3.59\pm0.04$ & $-1.062\pm0.005$ & $-1.657\pm0.003$ & $-2.451\pm0.009$ \\ \hline
    \end{tabular}
\end{table*}

\begin{figure*}
	\centering
	\includegraphics[scale=0.59]{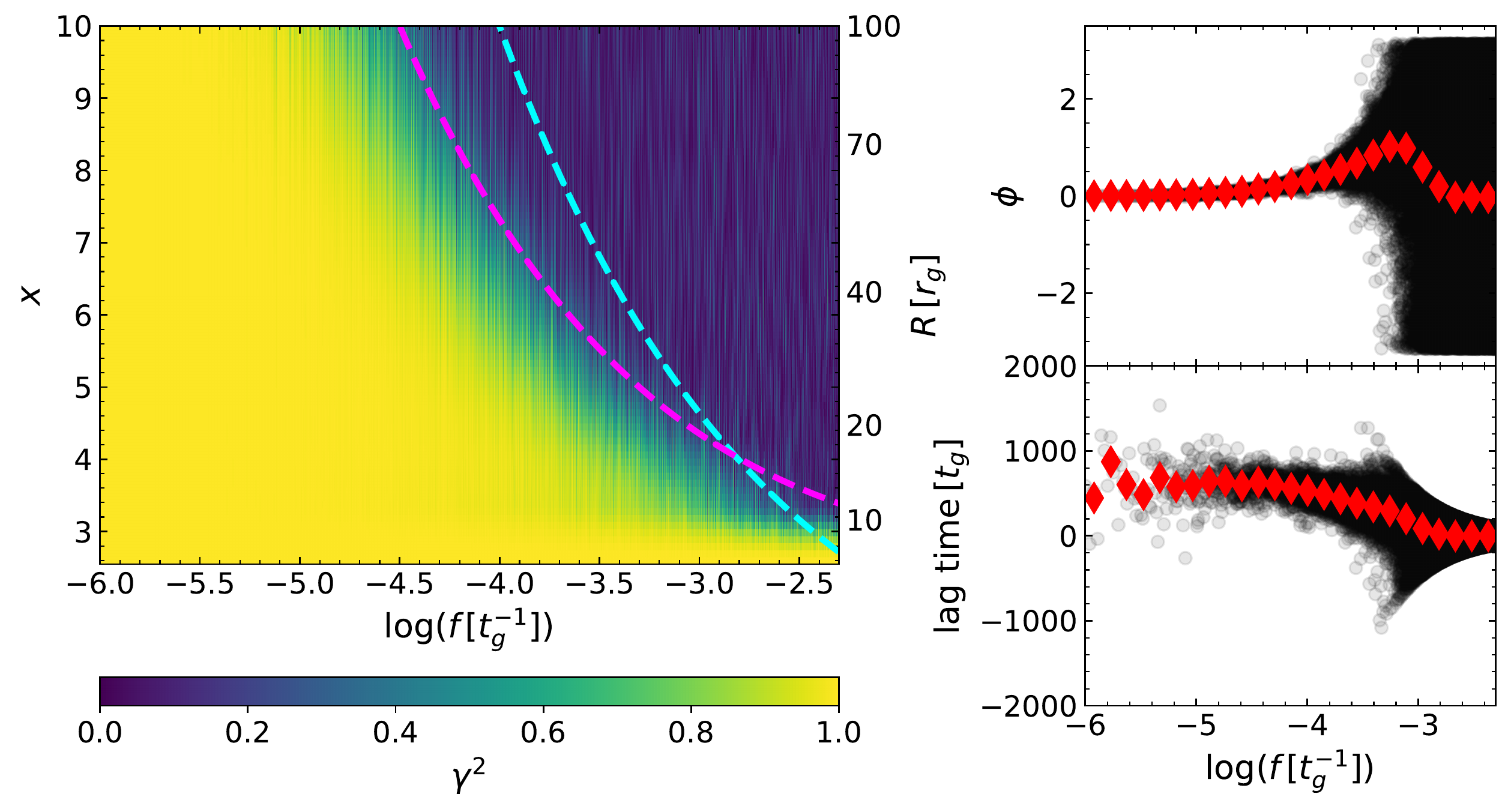}
	\caption{As for Fig. \ref{fig:coherence_map} but using $\Hcal=0.3$ rather than $0.1$.}
	\label{fig:coherence_map_03}
\end{figure*}

{ An opposite effect can be seen in the accretion rate PSDs in Table \ref{tab:aspect_PSD}. Here, larger values of $\Hcal$ give rise to larger values of $f_\text{break}$. While the driving timescale is independent of $\Hcal$, the viscous travel time is not and larger values of $\Hcal$ give shorter viscous travel times. We would expect these shorter travel times to increase coherence on the fastest timescales and thus increase high frequency power. The decreasing value of $f_\text{break}$ shows this increase in high frequency power and so is good evidence to support the explanation of the accretion rate PSD break frequency in $\S$\ref{sec:Fid_Res}.

We are now in a position to check that the larger values of $\Hcal$ do increase coherence in the disc.} In Fig. \ref{fig:coherence_map} we showed that the change from coherence to incoherence matches the travel time between radii. In order to check whether this agreement is a general property of our model, Fig. \ref{fig:coherence_map_03} shows the same coherence map as Fig. \ref{fig:coherence_map}, but with $\Hcal=0.3$. Comparison between the two coherence plots shows that the dissipation is in general more coherent when $\Hcal=0.3$. The viscous travel time is, as before, a good guide to where the change from coherence to incoherence occurs { and so overall coherence has increased through the disc}. This is potentially untrue at small radii, where the dissipation is seen to be less coherent than might be predicted by the travel time.

Fig. \ref{fig:coherence_map_03} also shows the phase and lag time between $x=5$ and $x=\sqrt{6}$. Where the dissipation is coherent, the lag time appears to be roughly constant at $\sim600t_g$. As for $\Hcal=0.1$, this is less than the viscous travel time of $2000t_g$. However, the lag time has decreased by a factor of $\sim9$ compared to $\Hcal=0.1$, the same factor as the {viscous} travel time.

\subsection{Driving Timescale}\label{sec:timescale}

We can now look at the effect of changing the driving timescale. In these models we use ${\alpha_0=\Hcal=0.1}$, and so the physical timescales described in $\S$\ref{sec:stoch_model} are related through ${\tilde{t}_{\nu,\text{g}} = 100\tilde{t}_{\nu,\text{c}} = 1000\tilde{t}_\phi}$. In this section we explore models with driving timescales between $\tilde{t}_\phi$ and $\tilde{t}_{\nu,\text{g}}$. In all of the models, the luminosity and instantaneous accretion rate are normalised to have a mean of unity. We use the standard deviation to quantify the width of the distributions. Since they have a mean of unity, we are essentially calculating $\sigma_L/\langle L\rangle$ and similar for $\dot{M}$.

\begin{figure}
	\centering
	\includegraphics[width=\columnwidth]{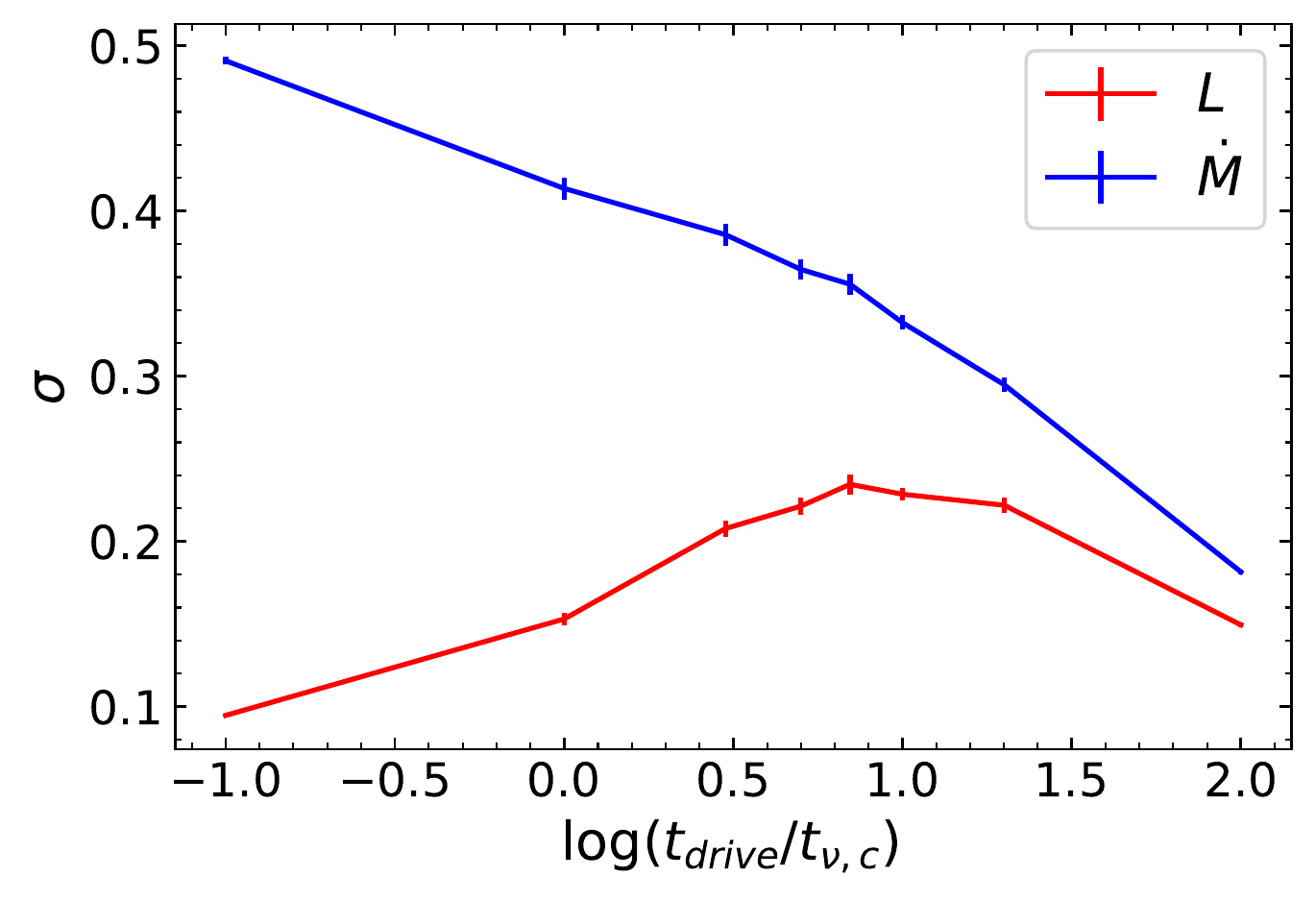}
	\caption{The standard deviation of the accretion rate across the ISCO (blue) and the bolometric luminosity (red) as a function of the driving timescale, expressed as multiples of the coherence length viscous timescale.}
	\label{fig:timescale_sigma}
\end{figure}

These results are shown in Fig. \ref{fig:timescale_sigma}. We can see that, while the $\dot{M}$ curve decreases with the driving timescale, the luminosity peaks at around $8\tilde{t}_{\nu,\text{c}}$. To understand this, when $\Hcal$ is constant, the diffusion equation (eq. \ref{eq:diffusion_eq_x}) acts to smooth out perturbations in $\alpha\Psi$. In our model, fluctuations in $\alpha$ are driven and so $\Psi$ responds to these perturbations. As we can see in eq. \eqref{eq:accretion_rate} with constant $\Hcal$, $\dot{M}$ depends on the {radial} derivative of $\alpha\Psi$ calculated at the inner edge of the disc.

We know that in steady state, the accretion rate will be constant. If $\alpha$ varies very slowly (long driving timescale), then the surface density (${\Psi/x^2}$) will be able to respond effectively to the changes in $\alpha$ and so the fluctuation in $\dot{M}$ will be relatively small. Since the steady state solutions do not depend on $\nu(R)$, if the driving timescale is sufficiently long, the surface density will always be in steady state for the current viscosity. Therefore, in the limit that ${\tilde{t}_\text{drive}\rightarrow\infty}$ then we would expect ${\sigma_{\dot{M}}\rightarrow0}$.

At the other extreme, if $\alpha$ varies very quickly then $\Psi$ is unable to respond, and the variation in $\dot{M}$ is essentially given by the instantaneous variation of $\alpha$. This model uses the exponential model for $\alpha$ with ${\sqrt{\left<\beta^2\right>}=0.5}$ which gives a standard deviation for $\alpha$ of ${\sigma_\alpha=0.60}$. We would expect this to provide an upper limit for $\sigma_{\dot{M}}$, which is consistent with the results in Fig. \ref{fig:timescale_sigma}.

The crossover between these two extremes will depend on the timescale for $\Psi$ to respond to deviations away from steady state. Calling this response timescale $\tilde{t}_\text{res}$, the diffusion equation (eq. \ref{eq:diffusion_eq_x}) can be approximated as
\begin{equation}
    \label{eq:diffusion_response}
    \frac{\Psi}{\tilde{t}_\text{res}} \sim \frac{3}{4x}\frac{\alpha\Hcal^2\Psi}{(\Delta x)^2} \, ,
\end{equation}
where $\Delta x$ quantifies the characteristic length scale for perturbations. In defining a coherence length for $\beta$ (eq. \ref{eq:coherence_length}) we have introduced this length scale. Using this for $\Delta x$ gives the response timescale as
\begin{equation}
    \label{eq:response_timescale}
    \tilde{t}_\text{res} \sim \frac{x^3}{3\alpha_0} = \frac{1}{3}\tilde{t}_{\nu,\text{c}} \, ,
\end{equation}
where we have also taken $\alpha\sim\alpha_0$.

Returning to Fig. \ref{fig:timescale_sigma}, the plot for the luminosity curve is more complicated, showing maximum variability at an intermediate timescale. Here, the luminosity depends on the integral of $\alpha\Psi$ as opposed to its derivative at one location as in the case of $\dot{M}$. At long driving timescales, the same analysis as for the accretion rate can be applied. Here, $\Psi$ has sufficient time to respond to $\alpha$ fluctuations and so the luminosity is close to steady state at all times.

On very short timescales, we might expect a similar process to occur as for $\dot{M}$. Since $\alpha$ varies very quickly, we would expect $\Psi$ to be approximately constant, and so the local dissipation rate to be given simply by the fluctuations in $\alpha$. While this may be true, the luminosity is an integrated property rather than a local property as for $\dot{M}$ and so fluctuations {cancel} out. We propose that this is the reason why the variability of the luminosity is so much smaller than the variability of the accretion rate for small driving timescales.

Thus far, this discussion has ignored the effect of any propagation of the fluctuations. If the driving timescale is so short that it does {not} create any significant fluctuations in $\Psi$, then there can be only negligible propagation and combination of these fluctuations. We therefore propose that this is the cause of the peak in the variability at intermediate timescales. Here, the driving timescale is not so long that the disc is in pseudo-steady state, but is long enough to create sufficiently large fluctuations in $\Psi$ which can propagate and combine to create large luminosity fluctuations. We might expect that this peak should occur at $\tilde{t}_\text{res}$, the response timescale of the disc. However, examination of Fig. \ref{fig:timescale_sigma} shows that the luminosity variability peaks around $8\tilde{t}_{\nu,\text{c}}$, $24$ times greater than the response timescale. The physical reason behind this peak timescale is not yet understood and requires more examination.

\begin{table*}
    \centering
    \caption{Parameters for the best fit broken power-laws to the PSDs given by simulations with different driving timescales. In performing these fits, the high frequency data was excluded. The cut-off frequency detailed in $\S$\ref{sec:Fid_Res} is scaled by the same factor as the driving frequency. This ensures that the same proportion of the data is excluded in each fit.}
    \label{tab:timescale_PSD}
    \begin{tabular}{ccccc} \hline
        $\tilde{t}_\text{drive}$ & $m_1$ & $m_2$ & $\log(f_\text{break}/t_g^{-1})$ & $\log(\text{max}(1/\tilde{t}_\text{drive}))$ \\ \hline
        $\tilde{t}_\phi$ & $-0.676\pm0.016$ & $-1.552\pm0.009$ & $-2.483\pm0.016$ & $-1.17$ \\
        $\tilde{t}_{\nu,\text{c}}=10\tilde{t}_\phi$ & $-0.996\pm0.016$ & $-1.631\pm0.009$ & $-3.47\pm0.02$ & $-2.17$ \\
        $100\tilde{t}_\phi$ & $-1.143\pm0.018$ & $-1.675\pm0.008$ & $-4.59\pm0.03$ & $-3.17$ \\
        $\tilde{t}_{\nu,\text{g}}=1000\tilde{t}_\phi$ & $-1.11\pm0.02$ & $-1.406\pm0.008$ & $-5.74\pm0.07$ & $-4.17$ \\ \hline
    \end{tabular}
\end{table*}

\begin{figure*}
	\centering
	\includegraphics[scale=0.59]{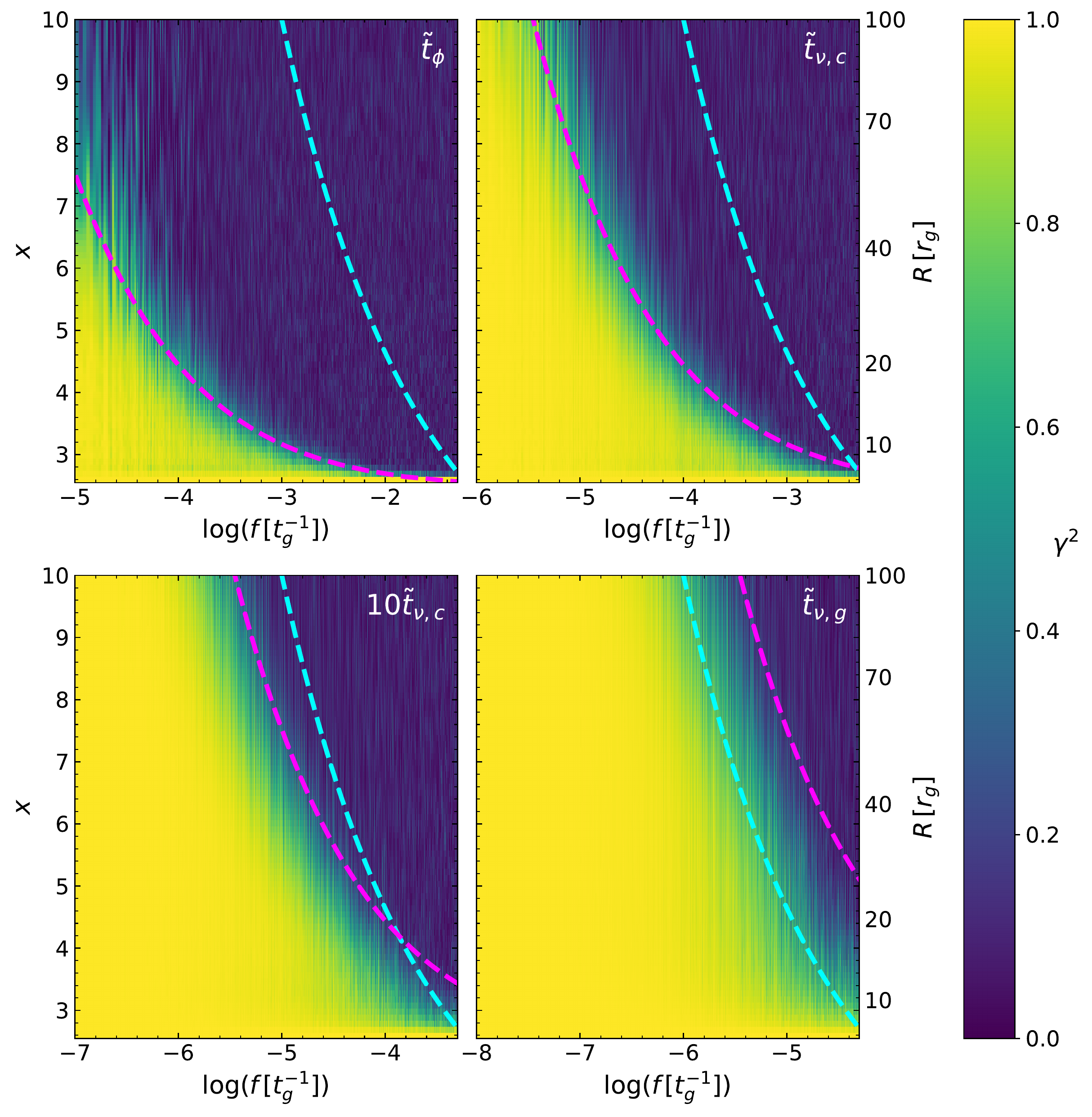}
	\caption{Coherence maps as shown in Fig. \ref{fig:coherence_map} but with different driving timescales. Note that the viscous travel time (magenta) is the same in each simulation but that the frequency range varies between simulations.}
	\label{fig:timescale_coh}
\end{figure*}

\begin{figure*}
	\centering
	\includegraphics[scale=0.57]{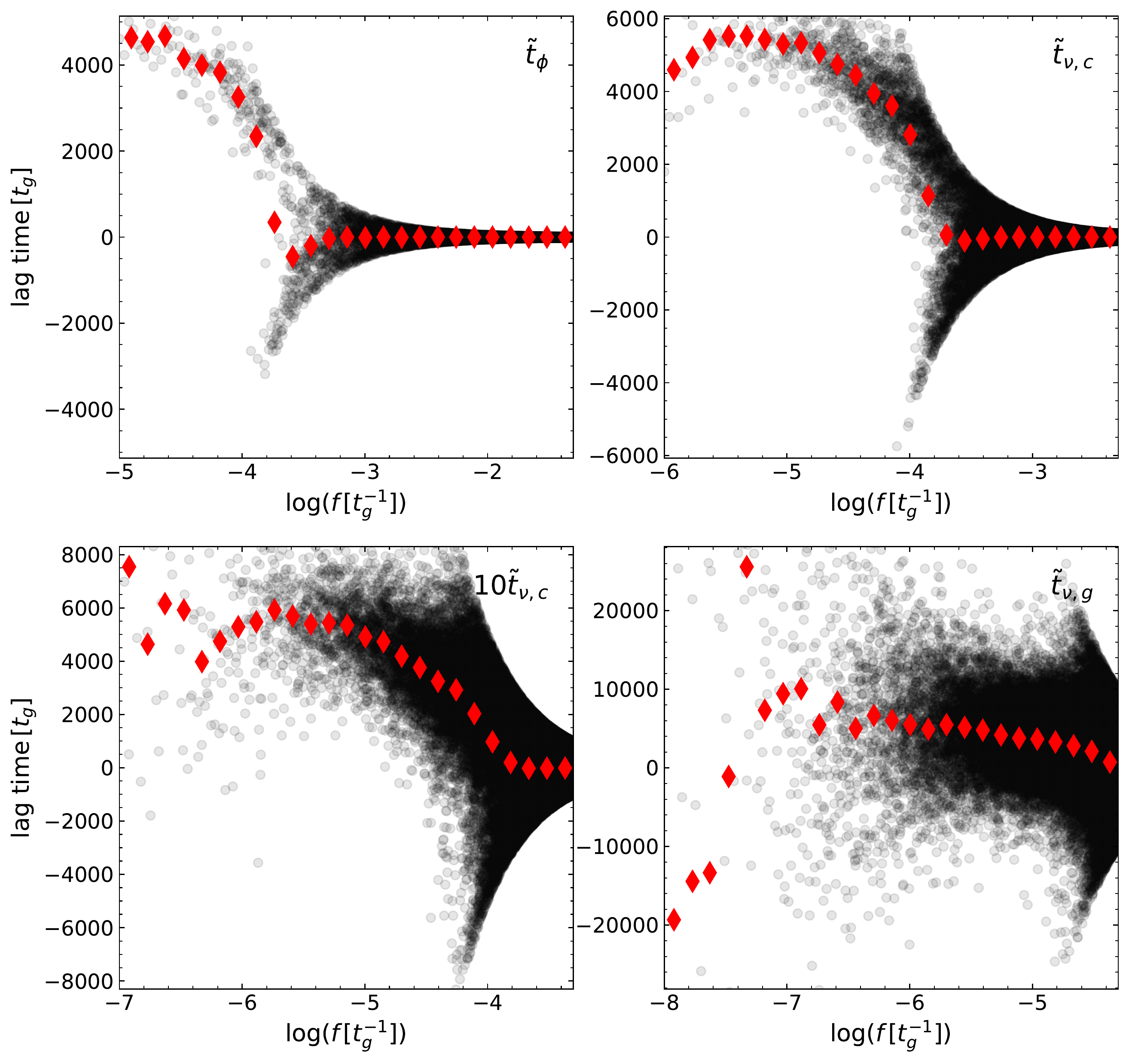}
	\caption{Time lags as shown in Fig. \ref{fig:coherence_map} but with different driving timescales.}
	\label{fig:timescale_phase}
\end{figure*}

Table \ref{tab:timescale_PSD} shows the best fit parameters for the broken power-law to the PSD (eq. \ref{eq:broken_powerlaw}). The low frequency slope, $m_1$, is theoretically expected to be flicker noise and so equal to $-1$ \citep{Lyubarskii1997}. In the simulation where the driving occurred on the orbital timescale, the slope is much less steep than expected. The slope is not dissimilar to that seen in the PSD for the accretion rate at the ISCO in the fiducial model. This slope suggests that the model with orbital timescale driving does not follow the standard paradigm of the model, possibly as they are not able to create the required propagation of fluctuations as suggested previously.  { However, we note that the luminosity from this orbital timescale model shows a strong preference for log-normal behaviour and a linear rms-flux relation, in the same way as the fiducial model (albeit with smaller rms variation as seen in Figure \ref{fig:timescale_sigma}).}

In our fiducial model, the accretion rate PSD in Fig. \ref{fig:PSD_Mdot} appeared to have three sections. The PSD of the accretion rate can also be calculated for the simulations with different driving timescales (not shown). The model with orbital timescale driving also shows these three sections. For models with ${\tilde{t}_\text{drive}>3\tilde{t}_{\nu,\text{c}}}$ the middle section becomes vanishingly small, suggesting that these models produce the expected results from the propagating fluctuations model. { The factor of 3 simply arises from looking at the accretion rate PSDs and is not obviously related to a physical parameter of the simulations.}

Looking at the comparison between the break frequency and the inverse of the shortest driving timescale, we can see that there is indeed a relation between them and that longer driving timescales give rise to {lower} break frequencies. As discussed previously, this break frequency is dependent on the emitting region of the disc and the frequency of the fluctuations at the outer edge of this region. In the same way as for eq. \eqref{eq:aspect_fit}, we can find a linear fitting formula between the driving timescale (expressed as a multiple of the orbital timescale) and the break frequency of
\begin{equation}
    \label{eq:timescales_fit}
    \log(f_\text{break}/t_g^{-1}) = m\log\left(\frac{t_\text{drive}}{t_\phi}\right) + c\, ,
\end{equation}
with ${m=-1.052\pm0.014}$ and ${c=-2.465\pm0.015}$.

In Fig. \ref{fig:timescale_coh} we show the coherence maps for simulations with four different driving timescales. In all of these models, the viscous travel time (eq. \ref{eq:SS_travel_time}) gives a good guide to the crossover between coherence and incoherence, including the simulation that was driven on the orbital timescale. This is perhaps surprising as it suggests that the propagating of fluctuations through the disc does occur in this simulation. This is in contrast to the low frequency slope of the PSD which shows that low frequency fluctuations from the outer edge of the disc are not transmitted to the inner regions as efficiently as would be expected.

{The propagation of fluctuations with orbital-timescale driving contrasts the findings of \citet{Cowperthwaite&Reynolds2014}.} They considered models where the driving occurred on both the orbital timescale $\tilde{t}_\phi$ and the global viscous timescale $\tilde{t}_{\nu,\text{g}}$. When they looked at the coherence in the dissipation between radii, they found that the simulation driven on the global viscous timescale showed similar coherence patterns to what we have shown previously. However, the simulation with driving on the orbital timescale was almost entirely incoherent. We attribute this to the prescription for the viscous fluctuations and, in particular, our enforcement of radial coherence on scales of $\Delta R=H$ in our current work. We have confirmed that, in the absence of such an enforcement \citep[as was the case in ][]{Cowperthwaite&Reynolds2014}, the local properties will be dominated by stochastic noise on unphysically small scales and any coherent propagations will be masked.

Finally, we can look at the effect the driving timescale has on the lag time, shown in Fig. \ref{fig:timescale_phase}. From this we can see that, in the coherent regions of the phase map (as shown in Fig. \ref{fig:timescale_coh}), the average time lag (as shown by the red diamonds) is approximately constant at $\sim5000t_g$. { However, it is clear that the two models with longer driving timescales show large amounts of scatter around that average lag time. To understand this, we need to consider the relationship between the driving timescale and the viscous travel time over the region between $x=5$ and $x=\sqrt{6}$. For the two longer driving timescale models, Fig. \ref{fig:timescale_coh} shows that the driving timescale (cyan) is comparable to or longer than the viscous travel timescale (magenta). This means that, as the value of $\alpha$ fluctuates slowly, the surface density in the disc has time to respond to these changes and reach some new steady state. This is the same effect as discussed in $\S$\ref{sec:timescale}. This adjustment of the surface density, local dissipation and local accretion rate to changes in viscosity at other locations in the disc requires the transfer of information both inwards and outwards through the disc. We propose that it is this transfer which results in the large range of lag times seen in Fig. \ref{fig:timescale_phase}. Despite this scatter, the average lag time is $\sim5000t_g$ and so the signature of the propagating fluctuations can still be seen. In the two simulations with driving timescales shorter than the viscous travel timescale (as seen in Fig. \ref{fig:timescale_coh}) the viscosity varies too rapidly for the disc to adjust into a new steady state. We therefore do not see the large scatter in lag time in these simulations.}

\section{Energy-resolved variability}
\label{sec:Energy}

\begin{figure}
	\centering
	\includegraphics[scale=0.57]{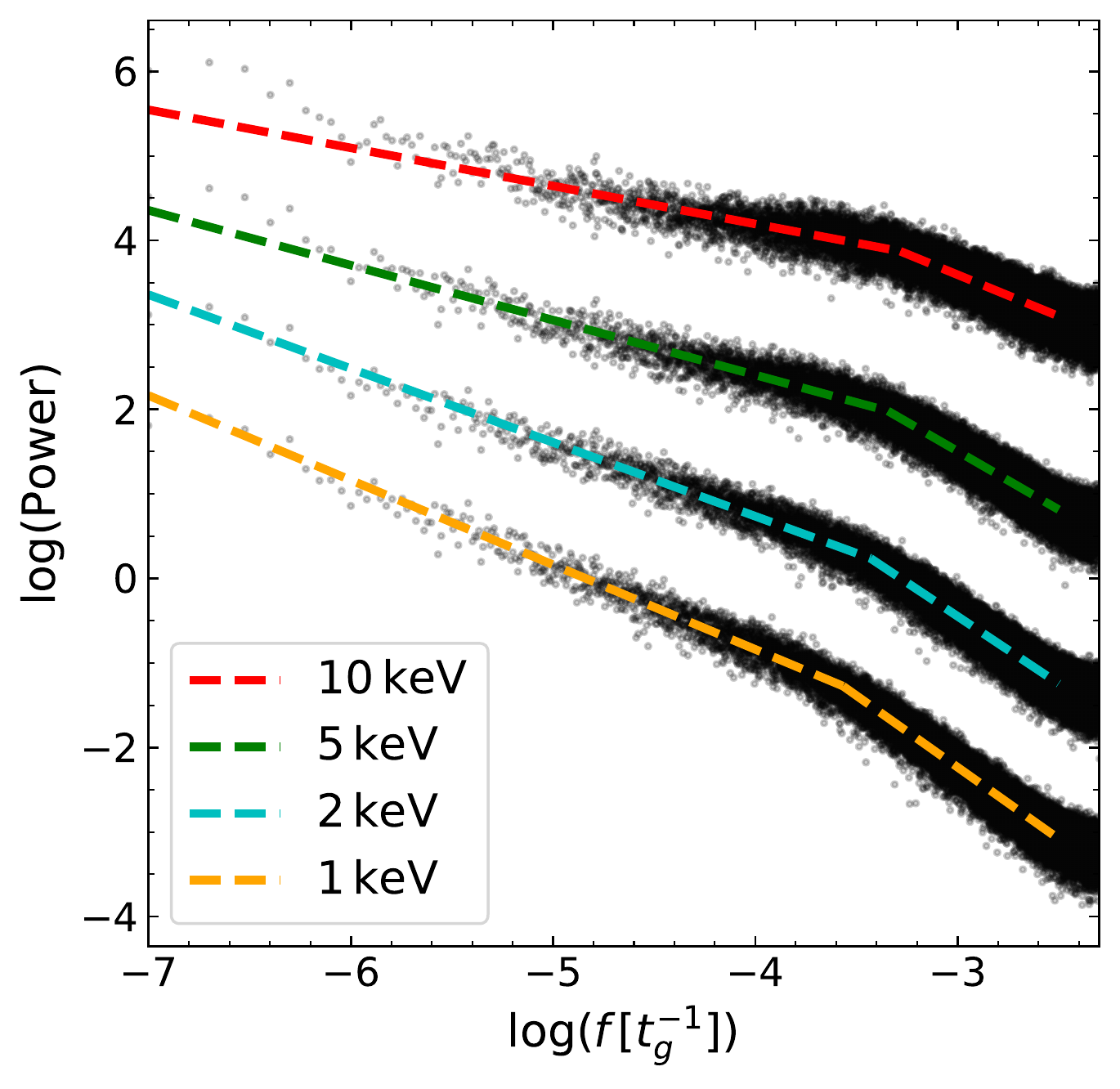}
	\caption{The PSD and best fit broken power-law model for the radiation from three different wavebands. The vertical offset of the PSDs is not indicative of the total power in each band and this has been adjusted to ensure the PSDs are ordered from highest to lowest energy band vertically.}
	\label{fig:band_PSD}
\end{figure}

Thus far, the analysis has involved the bolometric luminosity of the disc, or the total dissipation rate at any given radius. {Observational studies of coherence and frequency-dependent time-lags compare lightcurves drawn from different energy-bands, and we must recognize that even a monochromatic lightcurve has contributions from a range of radii in the disc and thus mixes together the radially-resolved information that we have been discussing.  Here, we briefly discuss energy-resolved results from our disc models.}

We assume that the disc radiates locally as a blackbody {with an effective temperature $T_{\rm eff}$ given by 
\begin{equation}
    \sigma_{\rm SB} T_{\rm eff}^4(R,t)=D(R,t).
\end{equation}
The monochromatic luminosity at energy $E$ and time $t$ is then given by integrating these multi-temperature Planck spectra across the disc radii,
\begin{equation}
    L(E,t)=\int \frac{E^3}{2\pi^2 \hbar^2c^2}\frac{1}{e^{E/kT_{\rm eff}(R,t)}-1}4\pi R\, {\rm d}R
\end{equation}
In this first exploration, we do not account for any distortions of the disc spectrum away from the blackbody form due to radiation transfer in the disc atmosphere, i.e., we adopt a colour correction factor of unity.

As an illustrative example, we adopt dimensional parameters appropriate to a black hole X-ray binary. We }use a black hole mass of ${M_\bullet=10M_\odot}$ and an accretion rate of ${\dot{M}=2\times10^{18}\,\text{g}\,\text{s}^{-1}}$, corresponding to an Eddington fraction of $0.12$ {assuming a non-spinning black hole.} For this system ${t_g=5\times10^{-5}\,\text{s}}$. We then analyse the lightcurves in four frequency bands, centered around $1$, $2$, $5$ and ${10\,\text{keV}}$. The width of each of the bands is $\pm10\%$ of the central value (i.e. ${0.9-1.1\,\text{keV}}$) for the lowest energy band. For this set of parameters, the maximum temperature of the steady state disc is ${4.6}$ million $\text{K}$. A single blackbody of this temperature peaks at an energy of ${1.1\,\text{keV}}$.

Once the light-curve for each of these three bands has been calculated, we can compute the PSDs for each of them in exactly the same way as for the bolometric luminosity. The three PSDs and their best fit power-laws are shown in Fig. \ref{fig:band_PSD} {with} best fit broken-power law parameters shown in Table \ref{tab:band_PSD}. We can see that the higher energy bands have shallow low frequency gradients and higher break frequencies.

Physically, the higher energy radiation originates from the hot, inner regions of the disc. In contrast, the lower energy {emissions originating} from a wider region of the disc. Some of the region that produces this lower energy radiation lies at moderate radii, and is not affected by the fluctuations created at smaller radii. This means that its break frequency is much lower as the highest frequency fluctuations (from the inner regions of the disc) do not affect this band. In contrast, the high energy bands produced in the inner regions are affected by fluctuations from almost all radii and so have a much higher break frequency.

\begin{table}
    \centering
    \caption{Parameters for the best fit broken power-laws for the PSDs shown in Fig. \ref{fig:band_PSD}.}
    \label{tab:band_PSD}
    \begin{tabular}{cccc} \hline
        $E_\text{band}\,[\text{keV}]$ & $m_1$ & $m_2$ & $\log(f_\text{break}/t_g^{-1})$ \\ \hline
        $10$ & $-0.442\pm0.015$ & $-0.977\pm0.013$ & $-3.29\pm0.03$ \\
        $5$ & $-0.646\pm0.014$ & $-1.381\pm0.011$ & $-3.346\pm0.017$ \\
        $2$ & $-0.877\pm0.015$ & $-1.594\pm0.010$ & $-3.430\pm0.019$ \\  
        $1$ & $-1.01\pm0.02$ & $-1.692\pm0.008$ & $-3.57\pm0.02$ \\  \hline
    \end{tabular}
\end{table}

The distinction is similar to that between the behaviour of the bolometric luminosity and the accretion rate across the ISCO, as seen in $\S$\ref{sec:Fid_Res}. Indeed, looking closely at the PSD for the highest energy band in Fig. \ref{fig:band_PSD}, there is a hint of the three section shape (with a flat middle section) seen in the PSD of the accretion rate.

\begin{figure}
	\centering
	\includegraphics[scale=0.57]{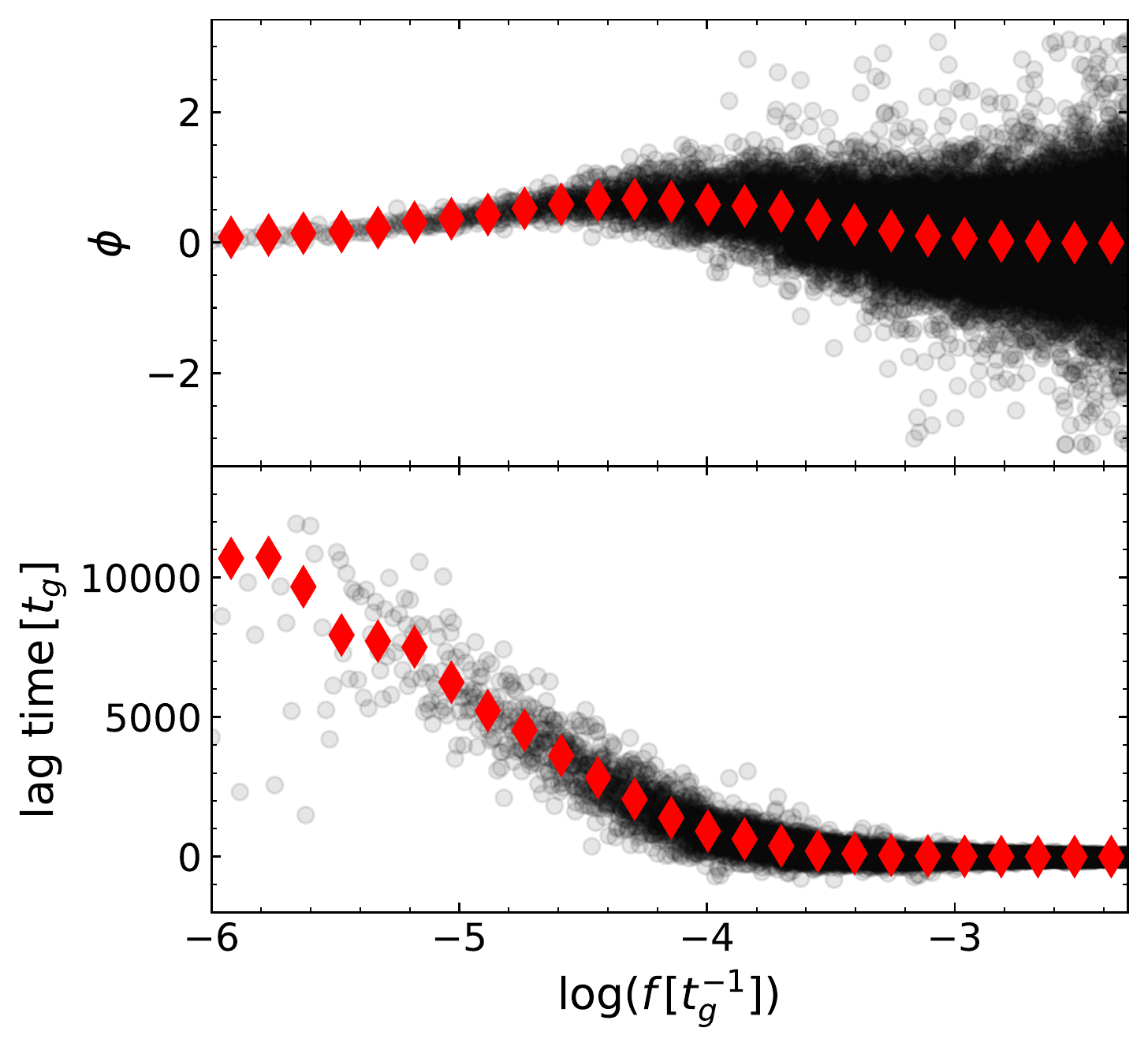}
	\caption{The phase (top) and time (bottom) lag between the ${5\,\text{keV}}$ and the ${1\,\text{keV}}$ energy bands. The black circles and red diamonds show unbinned and binned data respectively. Positive lags indicate that the lower energy band is leading the higher one.}
	\label{fig:band_lag}
\end{figure}

We can also calculate the phase and time lags between two energy bands. Figure \ref{fig:band_lag} shows the lag between the ${5\,\text{keV}}$ and the ${1\,\text{keV}}$ bands. The shape of the lags looks quite different to the lags between different radii (e.g. Fig. \ref{fig:coherence_map}), {highlighting the impact of combining the contributions from a range of radii}. The lag between different radii had an almost constant time lag whereas here the lag increases at lower frequencies. The phase lag tends to $0$ at both low and high frequencies, peaking at intermediate frequencies. { The lags between energy bands in observed sources show a strong dependence of the time lag on the frequency \citep{Arevalo+2006, Markowitz+2007, Lobban+2018} with lower frequencies being associated with larger lags, as produced by our simulations.}

\section{Discussion}

{This work expands upon the study of \citet{Cowperthwaite&Reynolds2014}, uncovering the dependence of the variability properties on the basic model parameters, especially disc thickness and the driving timescale for the stochastic fluctuations.} While there are key differences highlighted by the different runs, all the simulations show the hallmarks expected by the theory of propagating fluctuations and similar in nature to observations of accreting sources of many types. In particular, the luminosity and accretion rate distributions were found to be approximately log-normal in shape. In addition to this, the models all produce linear rms-flux relationships (which is expected for log-normal light curves). Fourier analysis reveal PSDs which are well fit by a broken-power law and show coherence between both the accretion rate and dissipation at different radii, with phase and time lags between the two, as expected from fluctuations in the disc propagating inwards.  It is of particular interest to note that these behaviours appeared under models in which the stochastic fluctuations in the viscosity are normally distributed. This implies that the observed log-normality is a direct result of the nature of the diffusion equation and does not place many restrictions on the nature of the viscosity fluctuations.

{Given the ubiquity with which broken power law PSDs are observed in real accreting systems, the physical nature of the break frequency is particularly important.} \citet{Lyubarskii1997} proposed that the break frequency is set by the viscous timescale at the inner edge of the disc. {Our work finds} strong evidence that the value of the break frequency is more complicated than this. Comparing the bolometric luminosity with the accretion rate at the ISCO, the luminosity has a lower break frequency. The reason for this is the extended area from which the luminosity originates. This means the high frequency fluctuations that are created from the inner region of the disc are unable to affect the entirety of the emitting region. Therefore the bolometric luminosity does not capture these higher frequencies and so the break frequency is lower. This is potentially of interest for observable systems. In real systems, a large fraction of the radiation comes not from the disc but from a compact and highly variable corona (\citealt{Liang&Nolan1984, White+1988} and review by \citealt{Uttley+2014}). It is plausible (though by no means proven) that the { instantaneous} luminosity of the corona {is more closely related to the instantaneous accretion rate at the ISCO than it is to the disc bolometric luminosity.}  In this case, the emission from the corona and that from the disc would be expected to produce PSDs of different break frequencies. This difference in break frequencies can be extended to include the thermal radiation from the disc in different energy bands. These were also shown to have different break frequencies with higher energy radiation (which originates from the innermost regions of the disc) having higher break frequencies than lower energy radiation.

Considering the break frequency for the PSD of the bolometric luminosity in particular, there appear to be two key dependencies of its {value} on the model parameters. Firstly, longer driving timescales give rise to smaller break frequencies. Intuitively, we would expect the frequency of the fluctuations produced through the disc the depend heavily on the frequency of the underlying perturbations in the viscosity. Secondly, it was found that thicker discs have lower break frequencies. The thicker discs show greater variability and thus have a larger emitting area. This area covers larger radii and so is less sensitive to the high frequencies fluctuations, giving rise to a lower break frequency. While the driving timescales would also be expected to affect the PSD of the accretion rate, the disc thickness would not. We have quantified the effect of both the disc thickness and the driving timescale on the break frequencies in the fitting formulae given in eqs. \eqref{eq:aspect_fit} and \eqref{eq:timescales_fit}. Assuming that the dependence on the two parameters are independent we can fit for both dependencies simultaneously. In physical units, this gives
\begin{equation}
    \label{eq:fitting}
    \log(f_\text{th}\, [\text{Hz}]) = m_\Hcal\Hcal + m_t\log\left(\frac{t_\text{drive}}{t_\phi}\right) - \log\left(\frac{M_\bullet}{M_\odot}\right) + c\, ,
\end{equation}
where ${m_\Hcal=-0.21\pm0.08}$, ${m_t=-1.055\pm0.014}$ and $c=2.859\pm0.018$ and the dependence on the mass of the central object has been made explicit. Note that the values of the gradients are slightly different to those in the individual fits in eqs. \eqref{eq:aspect_fit} and \eqref{eq:timescales_fit} due to fitting both values simultaneously.

When comparing our model results to real observations, we must remember that our model is far simpler than real systems and so we should not expect it to match with observations perfectly. However, it is informative to make broad comparisons and to see what areas our model falls short in explaining real systems. \citet{Scaringi+2013} observed PSDs for two different CVs. These PSDs show clear break frequencies with a low frequency slope that appears broadly consistent with $-1$ and a high frequency slope consistent with $-2$. This value of $-2$ is an important one in the literature, partly because of the popularity of the damped random walk model of AGN variability proposed by \citet{Kelly+2009}. This produces a Lorentzian PSD which has a high frequency slope of $-2$ but a low frequency slope of $0$. It should be noted that this model simply aims to model to variability and does not model any of the underlying physical processes.

Nevertheless, the high frequency slopes produced by our simulations are noticeably shallower (maximum slopes of around $-1.6$) than those seen in observations. Of particular note are the AGN PSDs obtained by \textit{Kepler} \citep{Smith+2018}. They found high frequency slopes in the range ${-1.7<m_2<-3.4}$. Steep slopes from \textit{Kepler} AGN are also found by \citet{Mushotzky+2011} and are not generally consistent with the \citet{Kelly+2009} model of AGN variability \citep{Kasliwal+2015}. While some ground-based optical AGN observations also report steep slopes \citep[e.g.][]{Simm+2016, Caplar+2017}, other studies show no such issues \citep[e.g.][]{Kelly+2009, Zu+2013}. Here we should note that the recent report of a potential instrumental systematic in \textit{Kepler} AGN \citep{Moreno+2020} which should be borne in mind. However, these steeps slopes, if indeed they are real, imply that there is some physics missing from our model which results in much less high frequency power than predicted by our model.

While some of the \textit{Kepler} AGN show log-normal distributions, the majority do not and further none of the objects show linear rms-flux relations. This is in direct contradiction with the predictions from this work and the more general field of propagating fluctuations. However, the other major theory behind the optical variability of AGN is based on the reprocessing of X-ray photons from the corona. \citet{Smith+2018} find no evidence that the variability originates from X-ray reprocessing. It is therefore a currently open question as to where this variability comes from.

Six of the \cite{Smith+2018} AGN require break frequencies to best fit their PSDs. Of these, five have masses and so we can calculate the break frequency divided by ${t_g^{-1}}$. The five AGN have break frequencies in the range ${-4.5<\log(f_\text{break}/t_g^{-1})<-3.7}$. As noted above, the break frequency is a strong function of the driving timescale and could therefore be used to try to understand the physical mechanism underpinning the variability. Error estimates are not given by \cite{Smith+2018}, but these values are not dissimilar to our fiducial model value of ${-3.47\pm0.03}$. The driving timescale in this model was based on the MHD simulations of \citet{Hogg&Reynolds2016} and the similarity between our results and observations are therefore encouraging.

XRBs in the thermal (high/soft) state are perhaps the closest systems to our model and characteristic values for these systems were used in $\S$\ref{sec:Energy}. One notable feature of XRBs is that they show much less variability in the thermal state than in the low/hard state \citep[e.g.][]{McClintock&Remillard2006}. The reason for this is not well understood but our simulations suggest one possible explanation. In the thermal state, the system is well modelled by a thin $\alpha$-disc. However, in the low/hard state the disc is radiatively inefficient and the cooling advection dominated, resulting in a much thicker disc \citep{Abramowicz+1995, Narayan&Yi1995b}. In this work we have shown that thicker discs show larger variability than thinner ones, due to them having a greater number of independent regions. While this effect is not large in our simulations, it could potentially be greater in 2-d . If incoherence occurs in the $\phi$ direction then this would lead to a greater number of independent regions, thereby reducing the overall variability.

\section{Conclusions}

In this work we have presented a suite of simulations, created using a simple numerical model for stochastically driven accretion discs, building on the work of \citet{Cowperthwaite&Reynolds2014}. From these simulations, we are able to draw a number of conclusions:

\begin{itemize}
    \item The simulations support the paradigm of the propagating fluctuations model. All of our simulations show log-normality in both the accretion rate and the bolometric luminosity and a linear rms-flux relation. They also show broken power-law PSDs, frequency dependent coherence and phase lags between the accretion rate and dissipation between radii. This is all as predicted by the analytic theory of propagating fluctuations.
    \item The accretion rate is not a suitable proxy for the bolometric luminosity. Specifically, the fact that the luminosity is an integrated quantity means that it shows less variation that the accretion rate at the ISCO. In addition, the lag time between dissipation at different radii is {shorter} than the lag between the accretion rates at the same radii by a factor of $\sim4$.
    { \item The log-normality and associated linear rms-flux relation are not dependent on the shape of the underlying viscosity distribution. This implies that the log-normality is a direct result of the disc equations and that observations of log-normality do not place constraints on the shape of the viscosity distribution.}
    \item Changing the disc thickness primarily affects the magnitude of the variability of the disc luminosity. Thicker discs have greater coherence meaning that there are fewer independent components to the luminosity integral and thus more variability.
    \item The driving timescale of the viscosity perturbations has a large effect on the results of the simulations. This is of particular interest as the cause of the viscosity fluctuations and the timescale for their variability is still uncertain. Driving on very short or very long timescales produced less variability in the luminosity than driving on intermediate timescales. In particular, driving on an orbital timescale produced much of the standard behaviour of the propagating fluctuations model, but appeared unable to efficiently transport the low frequency fluctuations from the outer regions of the disc.
    \item The break frequency of the PSD is of great interest. {Its strong dependence on the driving timescale opens a potential observational route for probing the underlying physics of the viscosity fluctuations and we report a fitting formula for its dependence on both the driving timescale and the disc thickness in eq. \eqref{eq:fitting}.} The accretion rate (which could influence the coronal emission) shows a higher break frequency that the integrated thermal bolometric emission from the disc. A similar effect is seen when considering different energy bands with higher energy bands having higher break frequencies.
    \item Qualitatively, our model produces realistic hard lags for realistic energy bands for the thermal disc emission.
\end{itemize}

Although the models we have used are relatively basic in terms of the physics involved, they are able to make interesting conclusions as given above. Future directions require relaxing some of the assumptions that have been made in this work. The most promising possibilities would be to relax the thin disc assumptions and expand to include advection dominated flows or expanding the model to more than one dimension.

\section*{Acknowledgements}

{The authors would like thank the referee for their report which improved the quality and clarity of the work.} S.G.D.T thanks support from the UK Science and Technology Facilities Council (STFC) Postgraduate Studentship program. C.S.R. thanks the STFC for support under the New Applicant grant ST/R000867/1 and Consolidated Grant ST/S000623/1, as well as the European Research Council (ERC) for support under the European Union’s Horizon 2020 research and innovation programme (grant 834203). 

\section*{Data Availability}

The data underlying this article and the code from which it was generated will be shared upon reasonable request to the corresponding author.




\bibliographystyle{mnras}
\bibliography{references} 




\appendix


\bsp	
\label{lastpage}
\end{document}